\documentclass[reprint,aps]{revtex4-1}
\usepackage{bbm}
\usepackage{amsmath}
\usepackage{amssymb}

\usepackage{epsfig}
\usepackage{graphicx}
\usepackage{color}
\usepackage{mathrsfs}
\usepackage{amsfonts}
\usepackage[english]{babel}
\usepackage{chngcntr}

\begin{document}

\title{Quantum criticality in sub-Ohmic systems with three competing terms: beyond conventional spin-boson physics }

\author{Nengji Zhou}
\author{Yulong Shen}
\author{Zhe Sun}
\email[Corresponding author:~]{sunzhe@hznu.edu.cn}
\date{\today}
\affiliation{School of Physics, Hangzhou Normal University, Hangzhou 311121, China
}

\begin{abstract}
Quantum phase transitions (QPTs) in the spin-boson model with/without the rotating-wave approximation (RWA) are systematically investigated through variational calculations using a sub-Ohmic bath with high spectral density. Four cases involving different system-environment interactions are examined, where transition points and critical exponents are accurately determined across varying tunneling strengths. Contrary to prior work, a rich phase diagram is revealed in the tunneling-coupling plane even at the low spectral exponent $s<1/2$, with a novel U(1)-symmetric phase being identified. As coupling increases, a multi-stage QPT sequence arises for the tunneling $0<\Delta < \Delta^*=0.074(1)$, whereas a single transition occurs beyond this range. Furthermore, an odd-parity phase is found to emerge even under the positive tunneling, exhibiting distinct characteristics relative to the prototype model.
\end{abstract}

\maketitle

\section{Introduction}
Dissipative quantum impurity models are of central importance in modern physics \cite{leg87}. Among these, the spin-boson model (SBM) has emerged as a cornerstone for understanding open quantum systems \cite{wei07}. This model describes the interaction between a two-level quantum system, such as a spin-$1/2$ particle, and a bosonic environment comprising an infinite number of harmonic oscillators \cite{hur08,bre16}. Its wide-ranging applications extend across various disciplines, including condensed matter physics (e.g., quantum impurity problems) \cite{bul05,gul11}, quantum optics (light-matter interactions) \cite{scu97, ota11,liu21}, quantum chemistry (proton-transfer reactions) \cite{ost16,kip09,col09}, and biological processes (energy transfer in photosynthetic complexes) \cite{eng07,ish12,hue13}.

A defining feature of the SBM is its ability to exhibit quantum phase transitions (QPTs), particularly the localized-delocalized transition driven by the competition between quantum tunneling and dissipation \cite{hur08}, wherein the coupling between the system and environment is characterized by a gapless spectrum $J(\omega) \sim \omega^s$ \cite{leg87}. The nature of the phase transition is critically dependent on the value of the exponent $s$, leading to a rich phase diagram. Specifically, the system exhibits a second-order transition in the sub-Ohmic regime $(s<1)$ and a Kosterlitz-Thouless (KT) type transition at the Ohmic point $(s=1)$. In contrast, no phase transition occurs in the super-Ohmic regime $(s>1)$. In the strong system-environment coupling regime, the model also manifests as a coherence-incoherence dynamic transition \cite{nal13,wan16,dan21,ott22}, which has significant implications for quantum control and information processing. Over the last decades, extensive analytical and numerical studies have been carried out on the SBM, cementing its role as a paradigm for exploring open quantum system phenomena \cite{bul03,voj05,win09,chi11,guo12,ton12,zho14}. Recent advances include investigations into collective spin-boson systems \cite{win14,fil21,per23}, nonequilibrium quantum dynamics \cite{ott22,bon24,tak24,zha24}, and quantum algorithms for quantum computing \cite{mie21,lep25}.

Recent research has increasingly focused on extending the SBM to frustrated interaction architectures, motivated by their unique capacity to host exotic collective phenomena \cite{har23,cer24}. For instance, studies on the SBM with diagonal and off-diagonal coupling to two identical, independent baths have revealed a tripartite phase structure which includes a quantum critical phase, a delocalized free phase, and a symmetry-breaking localized phase, in the absence of bias and tunneling. A continuous QPT separating the localized phase from the critical phase has been claimed to exist only at the spectral exponent $s^{*} < s < 1$ \cite{guo12,zho15}. Recent breakthroughs have extended this formalism to the SU($2$)-symmetric three-bath SBM \cite{web23}. However, the nontrivial tunneling effect remains largely unexplored in such models.

Advances in modern technology have enabled quantum simulations of qubit-boson coupled systems in platforms such as superconducting quantum circuits, cold atoms, and trapped ion systems \cite{lep18,mag18,lem18,kat23}. These experiments provide direct avenues for investigating QPTs. In certain implementations, anisotropic coupling emerges naturally through the suppression of counter-rotating terms \cite{gri13,xie14}. Concurrently, theoretical understanding of anisotropic quantum open systems has been significantly advanced. Critical phenomena in anisotropic quantum Rabi and Dicke models have been systematically characterized, with their phase diagrams identified to contain three fundamental domains: a normal phase and two superradiant phases separated by a first-order phase line \cite{bak14,liu17b}. These phases converge symmetrically at quantum tricritical points (QuTP) through second-order transition lines. For the anisotropic spin-boson model (ASBM), an asymmetric phase structure has been established in the shallow sub-Ohmic regime $s>0.5$, which comprises even-parity delocalized (I), odd-parity delocalized (II), and localized (III) phases, with coexisting QuTP. In contrast, the phase diagram in the deep sub-Ohmic regime $s<0.5$ simplifies to a single QPT separating localized and delocalized states, similar to the conventional SBM \cite{wan19b,wan20}. However, the fundamental characteristics of the emergent odd-parity quantum phase remain poorly understood. Furthermore, the potential existence of complex phase structures in the deep sub-Ohmic case can not be completely ruled out because the case under week tunneling has not been considered.

Given the lack of analytical treatments, various powerful numerical tools have been employed to investigate QPTs in the SBM \cite{won08, alv09,win09,guo12}. Among these, the numerical variational method (NVM) \cite{ber14,wan16,blu17,zho18,zho23} based on systematic coherent-state decomposition stands out for its excellent accuracy and reliability in addressing ground-state phase transitions and quantum dynamics. However, a recent study reported the NVM's failure to achieve the true ground state of the ASBM relative to the variational matrix-product-state approach (VMPS) \cite{wan19b}. A possible reason is that NVM calculations yield exact results only in the continuum limit corresponding to a high-density spectrum \cite{zho21,zho22}, which is lacked in the aforementioned reference.

In this work, ground-state properties of the ASBM coupled to a sub-Ohmic bath with high spectral density are systematically investigated using the NVM, considering three competing interactions. Transition boundaries and critical exponents are accurately determined, and the nontrivial tunneling effect is identified. Even in the deep sub-Ohmic regime, a rich phase diagram is found within the parameter space defined by tunneling and spin-bath couplings, contrasting prior findings. Beyond conventional localized and delocalized phase, the free phase exhibiting $U(1)$ symmetry and odd-parity phase are revealed.

The rest of the paper is organized as follows: the model and numerical variational approach are described in Sec.~\ref{sec:mod}, along with principal observation indexes. Numerical results are presented in Sec.~\ref{sec:num}. Finally, the phase diagrams and critical behaviors are discussed in Sec.~\ref{sec:dis}, followed by conclusions in Sec.~\ref{sec:con}.

\section{Methodology}\label{sec:mod}
\subsection{Model}
In terms of the Pauli creation/annihilation operators $\sigma_\pm=(\sigma_z \pm i\sigma_y)/2$, the Hamiltonian of the ASBM reads
\begin{eqnarray}
\hat{H}  = && \frac{\varepsilon}{2}\sigma_z-\frac{\Delta}{2}\sigma_x + \sum_{k}\hbar\omega_k b_{k}^\dag b_{k}  + \sum_k \left[\lambda_k(b^\dag_{k}\sigma_- + b_{k}\sigma_+) \right. \nonumber \\
           && + \left. \gamma_k(b^\dag_{k}\sigma_+ + b_{k}\sigma_-) \right],
\label{Hamilton}
\end{eqnarray}
where $\varepsilon$ ($\Delta$) denotes the energy bias (tunneling amplitude), $b^\dag_k$ ($b_k$) represents the bosonic creation (annihilation) operator of the $k$-th bath mode with frequency $\omega_k$, $\sigma_x$ and $\sigma_z$ are  spin-$1/2$ operators, and  $\lambda_k$ and $\gamma_k$ signify coefficients of the rotating wave (RW) and counter-rotating wave (CRW) couplings, respectively. For simplicity, we set Planck's constant $\hbar =1$, rendering model parameters $\Delta, \varepsilon,\lambda_k$, and $\gamma_k$ dimensionless.

From a theoretical perspective, the influence of the heat bath is fully characterized by a continuous spectral density function, $ J(\omega) = \sum_k \eta_k^2 \delta(\omega - \omega_k) $, where $ \omega $ is the frequency of the bosonic bath, and $ \eta_k $ denotes the coupling strength between the quantum spin and the $ k $-th bath mode. In the infrared limit ($ \omega \to 0 $), the power-law behavior of $ J(\omega) $ becomes particularly significant. To capture the low-energy spectral details, the spectral density is commonly expressed as $ J(\omega) = 2\alpha \omega_c^{1-s} \omega^s $ for $ 0 < \omega < \omega_c $  \cite{bul05, voj05, zha10, zho14, blu17}, where $ \alpha $ is the dimensionless coupling constant, and $ \omega_c = 1 $ represents the high-frequency cutoff. With a coarse-grained treatment based on the Wilson energy mesh, the values of $\lambda_k, \gamma_k$, and  $\omega_k$ in Eq.~(\ref{Hamilton}) are derived after dividing the phonon frequency domain $[0, \omega_c]$ into $M$ intervals $[\Lambda_k, \Lambda_{k+1}]\omega_c$ ($k=0, 1, \ldots, M-1$),
\begin{equation}
\label{sbm1_dis}
\eta_k^2  =  \int^{\Lambda_{k+1}\omega_c}_{\Lambda_k\omega_c}dt J(t), \quad \omega_k  =  \eta_k^{-2} \int^{\Lambda_{k+1}\omega_c}_{\Lambda_k\omega_c}dtJ(t)t,
\end{equation}
where $\Lambda_k=\Lambda^{k-M}$ \cite{bul03,hur08,chi11,fre13}. In this work, the incorporation of a bosonic bath with sufficiently high spectral density is employed to ensure reliable ground-state characterization, as numerical convergence in ground-state energy calculations requires the logarithmic discretization parameter $\Lambda \rightarrow 1$ \cite{zho22} Considering computational constraints, we set $\Lambda=1.05$ in the following, which is much closer to unity in comparison with those used in previous numerical studies \cite{bul03,hur08,chi11,fre13,ber14b}.

The Hamiltonian in Eq.~(\ref{Hamilton}) can also be expressed as
\begin{eqnarray}
\label{Hamilton2}
\hat{H}  = && \frac{\varepsilon}{2}\sigma_z-\frac{\Delta}{2}\sigma_x + \sum_{k}\omega_k b_{k}^\dag b_{k}  + \frac{\sigma_z}{2}\sum_k(\lambda_k+\gamma_k)(b^\dag_{k}+b_k)  \nonumber \\
           && + \frac{i\sigma_y}{2}\sum_k(\gamma_k-\lambda_k)(b^\dag_{k}- b_{k}),
\end{eqnarray}
In the absence of the energy bias $\varepsilon=0$, there are three distinct interactions involving the Pauli operators $\sigma_x,\sigma_y$, and $\sigma_z$, corresponding to the tunneling term and bath-induced diagonal and off-diagonal coupling terms, respectively. They compete with each other, resulting in rich quantum transitions. For simplicity, four special cases are considered in this work,
\begin{equation}
\label{fourcase}
\left\{
   \begin{array}{lccl}
     1.\text{\quad Diagonal coupling case:} & & &\lambda_k=\gamma_k=\eta_k /2 ,  \\
     2.\text{\quad Off-diagonal coupling case:} & &&\lambda_k=-\gamma_k=\eta_k /2,  \\
     3.\text{\quad RW coupling case:} & &&\lambda_k=\eta_k, \gamma_k=0,  \\
     4.\text{\quad CRW coupling case:} & &&\lambda_k=0, \gamma_k=\eta_k.
   \end{array}
\right.
\end{equation}
It should be noted that the strength of the total interactions $\eta=|\gamma|+|\lambda|$ are invariant for all the four cases.

\subsection{Method}
The earliest variational treatment of the SBM employed the polaronic unitary transformation which was introduced by Silbey and Harris \cite{sil84}.  This framework was subsequently adapted to study ground-state phase transitions in the sub-Ohmic SBM \cite{chi11} via the variational polaron ansatz. Further improvements were achieved  by superposing multiple coherent states and relaxing the imposed symmetry constraints of ($f_{n,k} \equiv \pm g_{n,k}$) \cite{naz12,zhe15,flo15,he18}. Numerical variational method was then formulated based on a systematic coherent-state decomposition of the many-body ground state \cite{zho14,blu17}.

In the NVM, a multiple polaron ansatz \cite{zho14}, also termed as the ``Davydov multi-$D_1$ ansatz'', is employed as the trial wave function,
\begin{eqnarray}
\label{vmwave1}
|\Psi \rangle & = & | \uparrow \rangle \sum_{n=1}^{N} A_n \exp\left[ \sum_{k=1}^{M}\left(f_{n,k}b_k^{\dag} - \mbox{H}.\mbox{c}.\right)\right] |0\rangle_{\textrm{b}} \nonumber \\
              & + & |\downarrow \rangle \sum_{n=1}^{N} B_n \exp\left[ \sum_{k=1}^{M}\left(g_{n,k}b_k^{\dag} - \mbox{H}.\mbox{c}.\right)\right] |0\rangle_{\textrm{b}},
\end{eqnarray}
where H$.$c$.\!$ denotes Hermitian conjugate, $| \uparrow \rangle$ ($| \downarrow \rangle$) represents the spin-up (spin-down) state, and $|0\rangle_{\rm b}$ is the vacuum state of the bosonic bath. In fact, Eq.~(\ref{vmwave1}) describes a superposition of the spin states that are correlated with the bath modes with displacements $f_{n,k}$ and $g_{n,k}$, where the subscripts $n$ and $k$ indicate the ranks. $M$ and $N$ represent the numbers of bath modes and coherent-superposition states, and $A_n$ and $B_n$ denote the weights of the coherent states. Notably, all variational parameters $f_{n,k}, g_{n,k}, A_n$, and $B_n$, are complex not real.

The ground state $|\Psi_{\rm g}\rangle$ is determined by minimizing the energy $E=\mathcal{H}/\mathcal{N}=\langle \Psi_{\rm g}|\hat{H}|\Psi_{\rm g}\rangle/\langle \Psi_{\rm g} |\Psi_{\rm g}\rangle$ where $\mathcal{H}$ and $\mathcal{N}$ represent the Hamiltonian expectation value and the normal of the wave function, respectively. A set of self-consistency equations is then derived using the Lagrange multiplier method,
\begin{equation}
\frac{\partial \mathcal{H}}{\partial x_{i}} - E\frac{\partial\mathcal{N}}{\partial x_{i}} = 0,
\label{vmit}
\end{equation}
where $x_i(i=1,2,3,\cdots, 2NM+2N)$ denotes any variational parameter. Finally, the iteration algorithmic is then adopted to numerically
solve self-consistency equations,
{\small
\begin{eqnarray}
A_n^{\rm next} & = &  \frac{\sum_{m}B_m\Gamma_{n,m}cc_{n,m}+\sum_m^{m\neq n}A_mF_{n,m}(aa_{n,m}-E)}{E-a_{n,n}}, \nonumber \\
B_n^{\rm next} & = &  \frac{\sum_{m}A_m K_{n,m}dd_{n,m}+\sum_m^{m\neq n}B_mG_{n,m}(bb_{n,m}-E)}{E-b_{n,n}},  \nonumber \\
f_{n,k}^{\rm next} & = & \frac{\sum_m^{m\neq n} A_mF_{n,m}\left[f_{m,k}(\omega_k+aa_{n,m}-E)+\frac{\lambda_k+\gamma_k}{2}\right]}{A_n(E-\omega_k-aa_{n,n})}  \nonumber \\
        & + &  \frac{\sum_mB_m\Gamma_{n,m}(g_{m,k}cc_{n,m} + \frac{\gamma_k-\lambda_k}{2})+ A_n\frac{\lambda_k+\gamma_k}{2}}{A_n(E-\omega_k-aa_{n,n})},   \\
g_{n,k}^{\rm next} & = & \frac{\sum_m^{m\neq n} B_mG_{n,m}\left[g_{m,k}(\omega_k+bb_{n,m}-E)-\frac{\lambda_k+\gamma_k}{2}\right]}{B_n(E-\omega_k-bb_{n,n})}  \nonumber \\
        & + &  \frac{\sum_mA_m K_{n,m}(f_{m,k}dd_{n,m} + \frac{\lambda_k-\gamma_k}{2})- B_n\frac{\lambda_k+\gamma_k}{2}}{B_n(E-\omega_k-bb_{n,n})}. \nonumber
\label{vmit2}
\end{eqnarray}
}
Where ${\rm aa}_{m,n}, {\rm bb}_{m,n}, {\rm cc}_{m,n}$, and ${\rm dd}_{m,n}$ denote
\begin{eqnarray}
aa_{m,n} & = & \sum_k\left[\omega_kf_{m,k}^{*}f_{n,k} + \frac{\lambda_k+\gamma_k}{2}(f_{m,k}^{*}+f_{n,k})\right]+\frac{\varepsilon}{2},   \nonumber \\
bb_{m,n} & = & \sum_k\left[\omega_kg_{m,k}^{*}g_{n,k} - \frac{\lambda_k+\gamma_k}{2}(g_{m,k}^{*}+g_{n,k})\right]-\frac{\varepsilon}{2}, \nonumber \\
cc_{m,n} & = & \sum_k\frac{\gamma_k-\lambda_k}{2}(f_{m,k}^{*} - g_{n,k})-\frac{\Delta}{2},  \\
dd_{m,n} & = & \sum_k\frac{\gamma_k-\lambda_k}{2}(f_{n,k} - g_{m,k}^{*})-\frac{\Delta}{2}, \nonumber
\label{vmfactor2}
\end{eqnarray}
respectively. Thus, the Hamiltonian expectation value and the normal of the wave function can be expressed as
\begin{eqnarray}
\mathcal{H} & = & \sum_{m,n}\left[ A_m^{*}A_nF_{m,n}aa_{m,n}+ B_m^{*}B_nG_{m,n}bb_{m,n} \right. \nonumber \\
            & + & \left. A_m^{*}B_n\Gamma_{m,n}cc_{m,n}+ B_m^{*}A_nK_{m,n}dd_{m,n} \right]. \\ \nonumber \\
\mathcal{N} & = & \sum_{m,n} \left[ A_m^{*}A_nF_{m,n} + B_m^{*}B_nG_{m,n} \right]. \nonumber
\end{eqnarray}
The Debye-Waller factors $F_{m,n}, G_{m,n},\Gamma_{m,n}$, and $K_{m,n}$ are defined as
\begin{eqnarray}
\ln F_{m,n} & = & \sum_{k}\left[f_{m,k}^{*}f_{n,k}-\frac{1}{2}(\|f_{m,k}\|^2+\|f_{n,k}\|^2)\right], \nonumber \\
\ln G_{m,n} & = & \sum_{k}\left[g_{m,k}^{*}g_{n,k}-\frac{1}{2}(\|g_{m,k}\|^2+\|g_{n,k}\|^2)\right], \nonumber \\
\ln \Gamma_{m,n} & = & \sum_{k}\left[f_{m,k}^{*}g_{n,k}-\frac{1}{2}(\|f_{m,k}\|^2+\|g_{n,k}\|^2)\right],  \\
\ln K_{m,n} & = & \sum_{k}\left[g_{m,k}^{*}f_{n,k}-\frac{1}{2}(\|g_{m,k}\|^2+\|f_{n,k}\|^2)\right], \nonumber
\label{vmfactor}
\end{eqnarray}
where $\|\cdots\|$ represents the modulus of the complex number. Using the relaxation iteration technique, one updates any variational parameter by the formula $x_i^{t+1}=x_i^{t}+f*(x_i^{\rm next}-x_i)$,
where the variable $x_i^{\rm next}$ is defined in Eq.~(\ref{vmit2}), and $f$ is the relaxation factor. In usual, $f=0.1$ is set in the variational procedure, while it gradually decreases to $0.001$ in the
simulated annealing algorithm.

In the Appendix~\ref{sec:appendix}, the convergence of the results is carefully assessed with respect to the number of the coherent superposition states ($N$) and  of effective bath modes ($M$). Results show that $M=430$ is sufficiently large for the convergence of NVM calculations when the logarithmic grid $\Lambda=1.05$ and lowest frequency $\omega_{\rm min} \approx 10^{-9}\omega_c$ are set \cite{zho22}. Limitations of computing resources, however, a modest number of coherent states, namely, $N=6$ for the diagonal coupling case and $N=4$ for the RW coupling case, is set in the following, which is sufficient for accurately depicting the phase diagram of the ASBM. The set of equations in Eq.~(\ref{vmit}) can be solved numerically using a global optimization algorithm, such as simulated annealing, with the relaxation iteration technique. To ensure convergence to the true ground state, in excess of a thousand random initial states are employed for each set of model parameters.

\subsection{Observables}
Using the normalized ground-state wave function $|\Psi_g\rangle$, we can determine the spin magnetization $|\langle\sigma_z\rangle|$ or $|\langle\sigma_y\rangle|$, the spin coherence $\langle\sigma_x\rangle$, and the von Neumann entropy $S_{\rm v-N}$ which characterizes the quantum entanglement between the spin and the surrounding bath,
\begin{eqnarray}
\label{entropy}
\langle\sigma_{x,y,z}\rangle & = & \langle\Psi_{\rm g}|\sigma_{x,y,z}|\Psi_{\rm g}\rangle, \nonumber \\
S_{\rm{v-N}} & = & -\rm Tr[\rho_{\rm s}log_2\rho_{\rm s}],
\end{eqnarray}
where $\rho_{\rm s}=\rm Tr_{\rm b}[\rho_{\rm sb}]$ is the reduced system density matrix obtained by tracing the total (system + bath) density operator $\rho_{\rm sb}$ over the bosonic bath.
To describe the ground-state properties of the bath, we introduce the variances of the phase space variables $\Delta X_{\rm b}$ and $\Delta P_{\rm b}$,
\begin{eqnarray}
\label{phase var}
\Delta X_{\rm b} & = & \langle \Psi_{\rm g}|(\hat{x}_k)^2 |\Psi_{\rm g}\rangle - \langle \Psi_{\rm g}|\hat{x}_k|\Psi_{\rm g}\rangle^2, \nonumber \\
\Delta P_{\rm b} &= &\langle \Psi_{\rm g}|(\hat{p}_k)^2 |\Psi_{\rm g}\rangle - \langle \Psi_{\rm g}|\hat{p}_k|\Psi_{\rm g}\rangle^2,
\end{eqnarray}
where the quadratures are defined as $\hat{x}_k = \left(b_k+b_k^{\dag}\right)/\sqrt{2}$ and $\hat{p}_k = i\left(b_k^{\dag}-b_k\right)/\sqrt{2}$. The departure from minimum uncertainty $QF(\omega_k)=\Delta X_{\rm b} \Delta P_{\rm b}- 1/4$ is then obtained, reflecting the quantum fluctuation of the bosonic bath mode at the frequency $\omega_k$.

Subsequently, symmetry properties of the ground states are also investigated. At the vanishing bias $\varepsilon=0$, the model admits  $Z_2$ symmetry, with the parity operator serving as its generator,
\begin{equation}
\hat{\Pi}=\exp(i\pi\hat{N}_{\rm ex}),
\label{symmetry}
\end{equation}
where $\hat{N}_{\rm ex}=\sum_k(b^\dag_{k}b_k)+\sigma_+\sigma_-$ represents the total number of the excitations.  Since $\hat{\Pi}^2=\mathbb{I}$, the eigenvalues of the parity operator $\hat{\Pi}$ are $\pm 1$, corresponding to the even/odd parity, respectively. The symmetry parameter $\langle\hat{\Pi}\rangle$ undergoes a sudden transition from unity to zero, signaling spontaneous symmetry breaking. This sharp jump is regarded as a dependable indicator for detecting QPTs \cite{wan19b,wan20,zho21}. Besides, the system possesses a $U(1)$ symmetry if the ground-state is unchanged by the action of the operator ${\rm Re}(\theta)=\exp(i\theta\hat{N}_{\rm ex})$ where $\theta$ represents an arbitrary angle. Thus, the total excitation number $\langle \hat{N}_{\rm ex}\rangle=0$ is expected.

Furthermore, the structure of the wave function also provides valuable insight into distinguishing different phases of the ASBM. Although variational parameters are complex, we introduce, for simplicity, their real, averaged counterparts, including the coherent-state weights $\overline{A}$ and $\overline{B}$, and the displacement coefficients $\overline{f}_k$ and $\overline{g}_k$,
\begin{eqnarray}
\label{vmamp}
\overline A & = & \pm \sqrt{\sum_{m,n}A_m^{*}A_nF_{m,n}}, \nonumber \\
\overline B & = & \pm \sqrt{\sum_{m,n}B_m^{*}B_nG_{m,n}},   \\
\overline{f}_k & = & {\rm Re}\sum_{m,n}\frac{A_m^{*}A_nF_{m,n}(f_{m,k}^{*}+f_{n,k})}{2\overline{A}^2}, \nonumber \\
\overline{g}_k & = & {\rm Re}\sum_{m,n}\frac{B_m^{*}B_nG_{m,n}(g_{m,k}^{*}+g_{n,k})}{2\overline{B}^2}.    \nonumber
\end{eqnarray}
Without loss of generality, we set $\overline{A} > 0$. Thus, $\overline{B}$ has the same sign as the spin coherence since $\langle\sigma_x\rangle=2\sum_{m,n}A_mB_n\Gamma_{m,n}$.

\section{Numerical results}\label{sec:num}

\begin{figure*}[htb]
\centering
\epsfysize=6cm \epsfclipoff \fboxsep=0pt
\setlength{\unitlength}{1.cm}
\begin{picture}(13,6)(0,0)
\put(0.0,0.0){{\epsffile{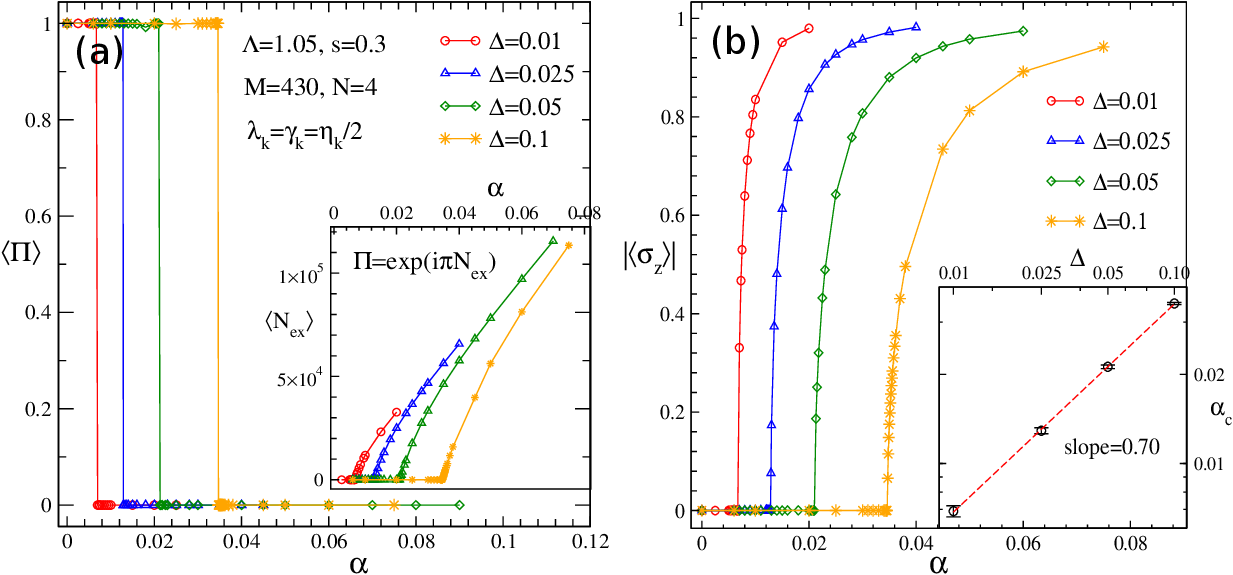}}}
\end{picture}
\caption{In the diagonal coupling case, the symmetry parameter $\langle\hat{\Pi}\rangle$ in (a) and the spin magnetization $|\langle\sigma_z\rangle|$ in (b) are plotted against the coupling strength $\alpha$ for different values of the tunneling amplitude $\Delta$. The insets show the total number of the excitations $\langle \hat{N}_{\rm ex} \rangle$  and the transition points $\alpha_c$, and dashed line represents a power-law fit. }
\vspace{0.5\baselineskip}
\label{f1}
\end{figure*}

\begin{figure*}[htb]
\centering
\epsfysize=6cm \epsfclipoff \fboxsep=0pt
\setlength{\unitlength}{1.cm}
\begin{picture}(13,6)(0,0)
\put(0.0,0.0){{\epsffile{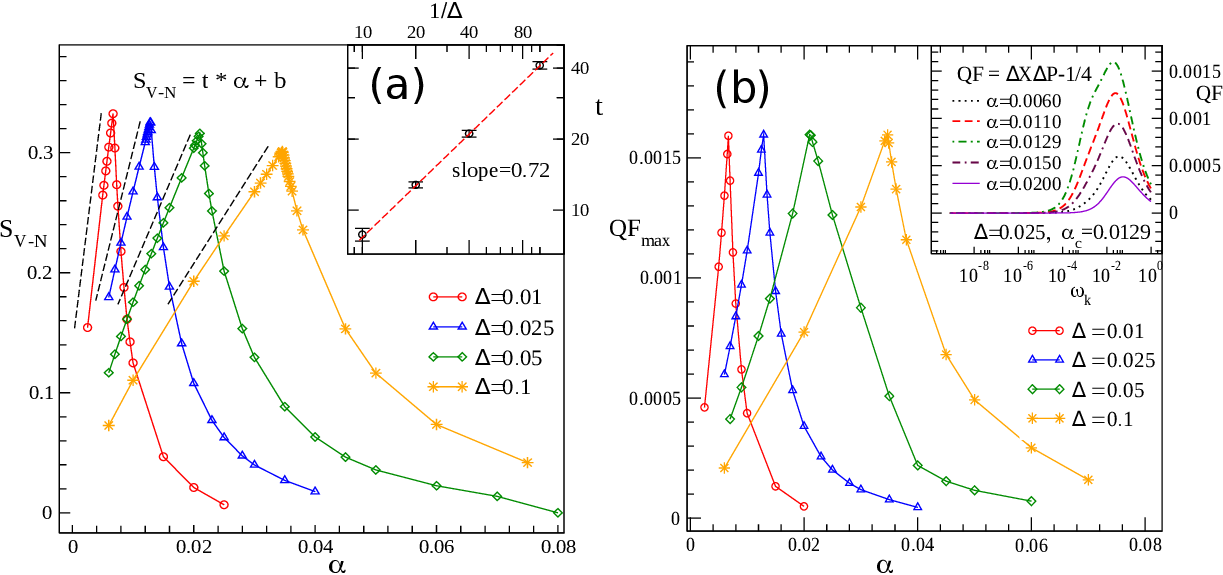}}}
\end{picture}
\caption{Quantum entanglement ($S_{\rm v-N}$) and quantum fluctuation ($QF = \Delta X_b\Delta P_b-1/4$) are presented as a function of the coupling strength $\alpha$ for different tunneling $\Delta$. In the insets, the slope of the curves $S_{\rm v-N}$ as well as the frequency-dependent $QF(\omega_k)$ for different coupling $\alpha$ are shown. Besides, dashed lines represent linear and power-law fits in (a) and its inset, respectively. In (b), $QF_{\rm max}$ denotes the peak value of the curve $QF(\omega_k)$. }
\label{f2}
\end{figure*}

With numerical variational calculations, ground-state properties of the anisotropic model embedded in sub-Ohmic Bath with a high dense spectrum are comprehensively studied for different values of the tunneling amplitude $\Delta=0.1,0.05,0.025,0.01$, and $0$, taking the setting of the spectral exponent $s=0.3$ as an example. The energy bias $\varepsilon=0$ is set unless otherwise noted. Statistical errors of the critical couplings and exponents are estimated by dividing the total samples into two subgroups. If the fluctuation in the curve is comparable with or larger than the statistical error, it will be taken into account. Four different cases mentioned in Eq.~(\ref{fourcase}) are taken as examples to investigate QPTs.

\subsection{ Diagonal coupling or off-diagonal coupling}

\begin{figure*}[htb]
\centering
\epsfysize=6cm \epsfclipoff \fboxsep=0pt
\setlength{\unitlength}{1.cm}
\begin{picture}(14,6)(0,0)
\put(0.0,0.0){{\epsffile{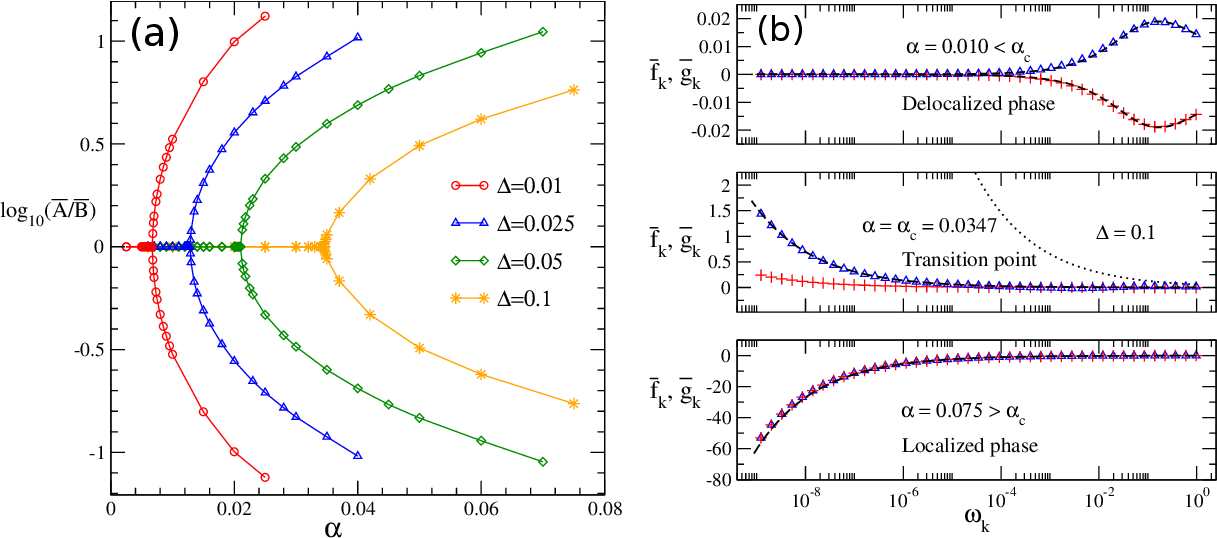}}}
\end{picture}
\caption{ (a) The logarithmic value of the ratio between two average coherent-state weights $\log_{10}(\overline{A}/\overline{B})$ is plotted with respect to the coupling $\alpha$ for different tunneling $\Delta$. (b) Average displacement coefficients $\overline f_k$ (pluses) and $\overline g_k$ (triangles) are displayed for the delocalized phase, the transition point, and the localized phase, taking $\alpha=0.010,~0.0347,$ and $0.075$ as examples. Dashed lines represent the fits with the optimal displacement formula. }
\vspace{0.5\baselineskip}
\label{f3}
\end{figure*}

\begin{figure*}[htb]
\centering
\epsfysize=11cm \epsfclipoff \fboxsep=0pt
\setlength{\unitlength}{1.cm}
\begin{picture}(13,11)(0,0)
\put(0.0,0.0){{\epsffile{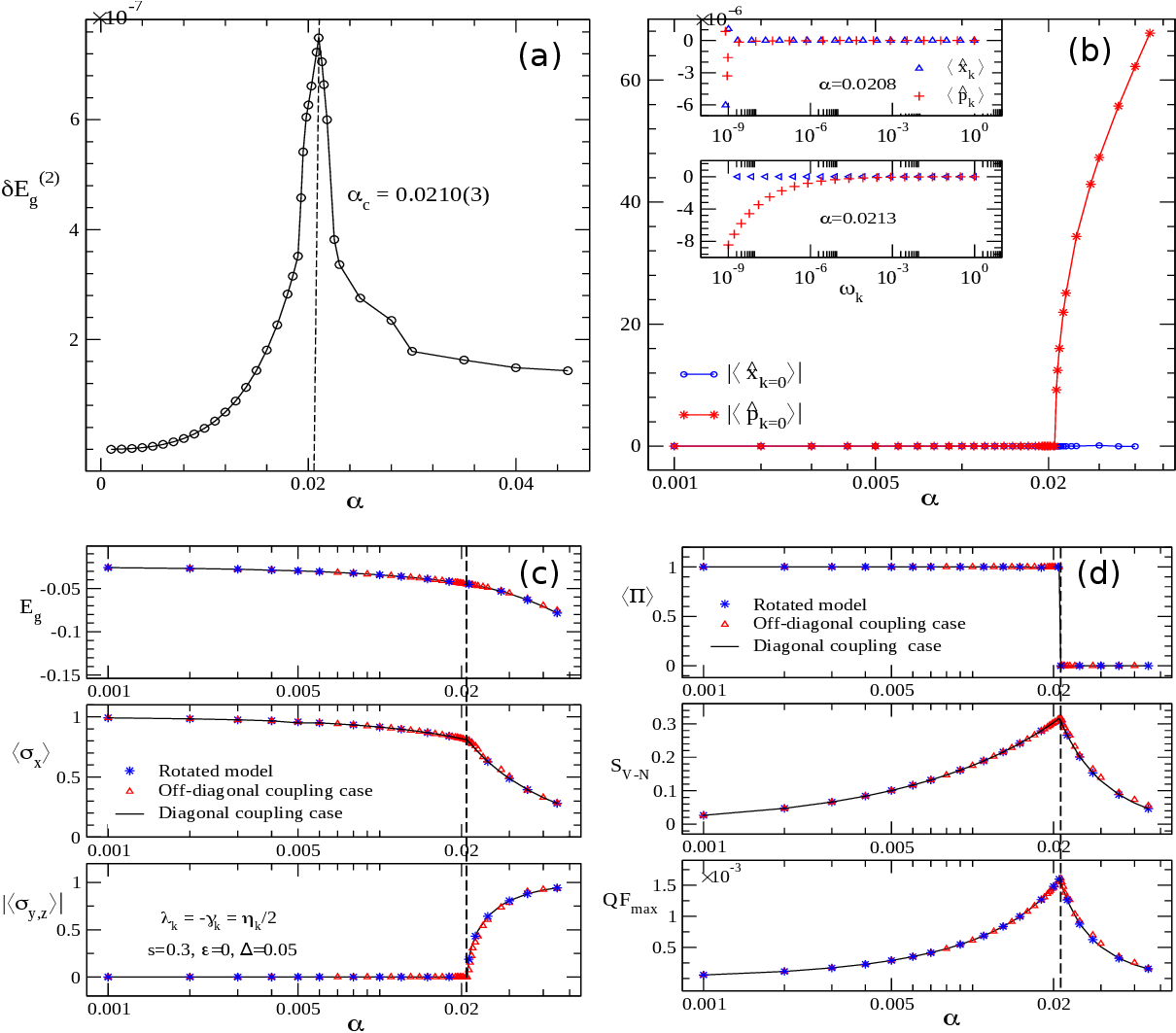}}}
\end{picture}
\caption{(a) In the off-diagonal coupling case under the tunneling $\Delta=0.05$, the variance of the ground-state energy $\delta E_g^{(2)}$ is displayed with respect to the coupling $\alpha$. The critical point $\alpha_c=0.0210(3)$ is marked by the dashed line. (b) Absolute values of the position $|\langle \hat{x}_{k=0}\rangle|$ and moment $|\langle \hat{p}_{k=0}\rangle|$ are plotted on a linear-log scale for the bath mode with the lowest frequency $\omega_{\rm min}$. In the inset, frequency-dependent behaviors are given for $\alpha=0.0208$ and $0.0213$. Besides, the ground-state energy $E_g$, spin coherence $\langle \sigma_x \rangle$, and spin magnetization $\langle \sigma_{y,z} \rangle$ in (c), and symmetry parameter $\langle\hat{\Pi}\rangle$ , von Neumann entropy $S_{\rm v-N}$, and maximum of the quantum fluctuation $QF_{\rm max}$ in (d) are shown, in comparison with those of the diagonal coupling case and rotated model. }
\label{f4}
\end{figure*}

In this subsection, the system-bath interaction type is set to diagonal ($\sigma_z$) or off-diagonal ($\sigma_y$) coupling.
According to the Hamiltonian in Eq.~(\ref{Hamilton2}), the anisotropic model reduces to the conventional SBM when $\lambda_k=\gamma_k=\eta_k /2$. As shown in Fig.~\ref{f1}(a), the symmetry parameter $\langle\hat{\Pi}\rangle$ defined in Eq.~(\ref{symmetry}) is displayed for various values of tunneling strengths $\Delta=0.01,0.025,0.05$, and $0.1$. Spontaneous symmetry breaking is confirmed by the discontinuity, with the critical point $\alpha_c$ estimated from the abrupt jump from $\langle\hat{\Pi}\rangle=1$ to $0$. In the inset, the total number of the excitations $\langle \hat{N}_{\rm ex} \rangle$ increases smoothly with the coupling $\alpha$ in the localized phase, but remains below unity in the delocalized phase. The behavior closely resembles that of the superradiant phase transition in the quantum Rabi model \cite{hwa15}. Figure~\ref{f1}(b) shows the order parameter $|\langle \sigma_z \rangle|$ versus $\alpha$ for different $\Delta$. Power-law scaling is expected with respect to the shift $\alpha-\alpha_c$. The critical coupling $\alpha_c$ is determined by minimizing the fitting error. These results show improved accuracy over subfigure (a). The dependence on the tunneling is presented in the inset with error bars, revealing a clear power law $\alpha_c(\Delta) \sim \Delta^{0.7}$ through numerical fitting. It agrees well with the analytical prediction $\alpha_c(\Delta) \sim \Delta^{1-s}$ \cite{chi11}.

Subsequently, higher-order statistics of QPTs are investigated, such as the quantum entanglement and quantum fluctuation, typified by the von Neumann entropy $S_{\rm v-N}$ and the departure from minimum uncertainty $QF$, respectively. In Fig.~\ref{f2}(a), the entanglement entropy $S_{\rm v-N}$ is presented in a range of the coupling $\alpha$ from $0$ to $0.08$ for different values of the tunneling $\Delta$. All the curves have cusp-like shapes, pointing to the occurrence of the singularity. It confirms that the phase transition is of second order. The slope $t$ characterizing the sharpness of the angle is then measured in the vicinity of the transition point $\alpha_c$, and results are presented in the inset. A pow-law increase of $t(1/\Delta)$ gives the exponent $0.72$, almost the same as that in Fig.~\ref{f1}(b), demonstrating the robustness of the $\Delta$-dependence.

In the localized phase ($\alpha > \alpha_c$), a linear dependence of the entanglement entropy on $\ln(\alpha-\alpha_c)$ is predicted by underlying quantum field theories \cite{cal04,bal12}. This behavior is expressed as
\begin{eqnarray}
\label{conformal field theory}
S_{\rm v-N} & \sim & A(c/6)\ln \xi \nonumber \\
&=& A(c/6)\ln k(\alpha-\alpha_c)^{-\nu}  \nonumber  \\
& = & b - A\nu(c/6)\ln(\alpha-\alpha_c)
\end{eqnarray}
where $\xi$ denotes the correlation length in imaginary time, $A$ represents the number of boundary points, $\nu$ is the correlation-length exponent, and $c$ is the central charge of the effective conformal field theory. The parameters $k$ and $b$ are defined as non-universal constants. This linear dependence is clearly observed in our numerical results (not shown), and a coefficient $Ac\nu/6 = 0.0812$ is extracted from the slope of the linear fit. Assuming $A=1$ and the mean-field exponent $\nu=1/2$, a central charge of $c \approx 0.97$ is estimated. As mentioned in Ref.~\cite{osh19}, the universal scaling of the inverse correlation length is found to be identical to that of the lowest energy gap, which was derived from energy-flow diagrams in NRG and VMPS methods \cite{bul03,guo12}, or by the lower bound of the critical frequency regime for the average displacement \cite{blu17}. Consequently, the value of the central charge can be obtained directly, without requiring the correlation-length exponent $\nu$ as an input. Furthermore, the relation between the left-right entanglement entropy, which reveals the system's intrinsic topological structure, and the central charge has been analyzed in previous analytical and numerical work \cite{das15,len19}. It is indicated that the linear dependence in Eq.~(\ref{conformal field theory}) may be universal.

In the insets of Fig.~\ref{f2}(b), quantum fluctuation $QF(\omega_k)$ is displayed at $\Delta=0.025$ for different couplings $\alpha=0.0060,0.0110,0.0129,0.0150$, and $0.0200$. Each curve exhibits a single peak in the high-frequency region, and the peak value reaches its maximum at the transition $\alpha_c$ (dash-dotted line). The peak value $QF_{\max}$ versus the coupling $\alpha$ is also displayed, and the behavior is quite similar to that of the von Neumann entropy $S_{\rm v-N}$.  The transition point identified via the singularity location agrees well with those obtained from the symmetry parameter, the total number of the excitations, and the spin magnetization.

\begin{figure*}[htbp]
\centering
\epsfysize=11cm \epsfclipoff \fboxsep=0pt
\setlength{\unitlength}{1.cm}
\begin{picture}(13,11)(0,0)
\put(0.0,0.0){{\epsffile{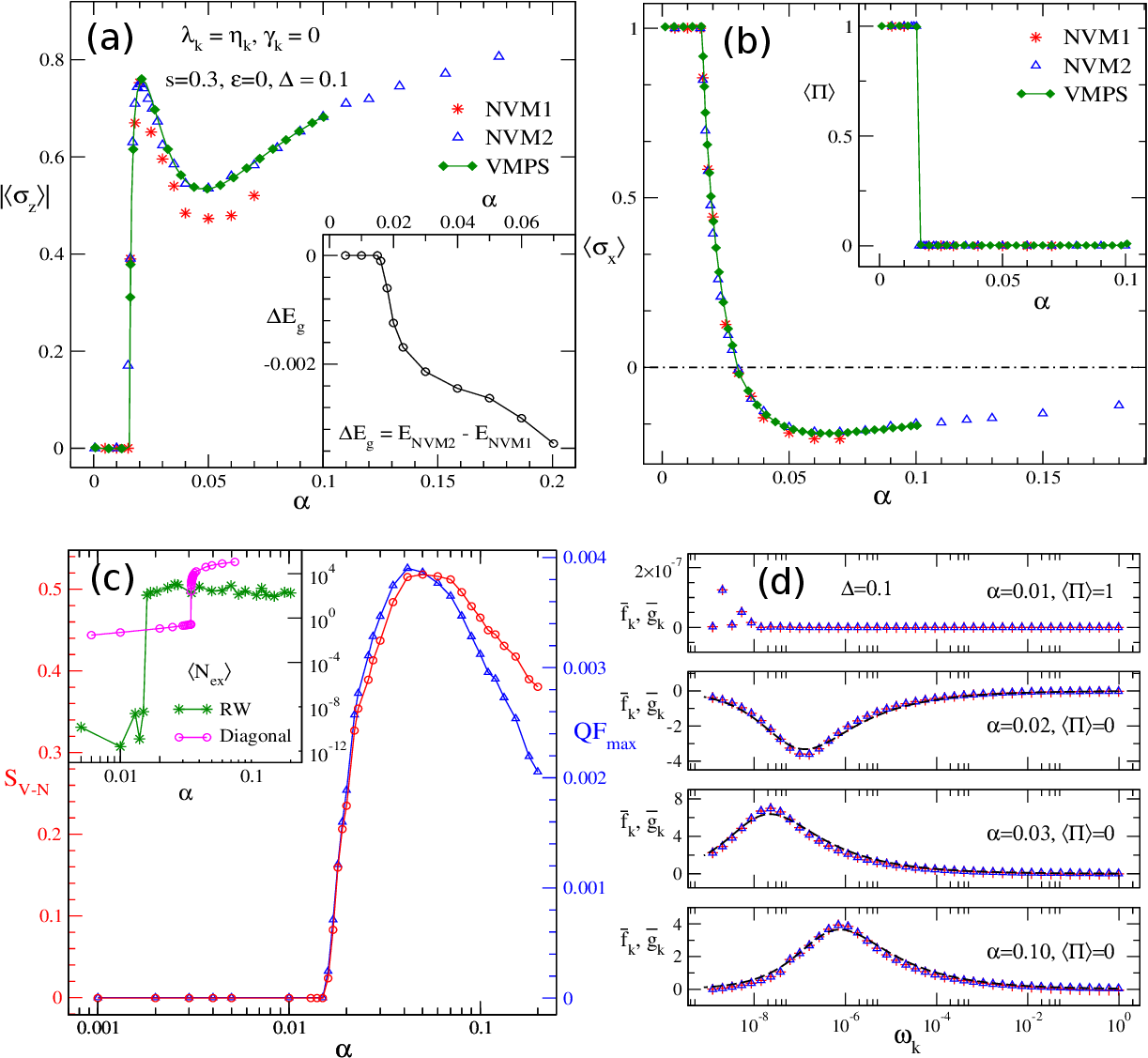}}}
\end{picture}
\caption{In the RW coupling case, numerical results obtained from NVM1, NVM2, and VMPS are presented for the spin magnetization $|\langle \sigma_z \rangle|$ in (a) and the spin coherence $\langle \sigma_x \rangle$ in (b), under the strong tunneling $\Delta=0.1$. The difference of the ground-state energies $\Delta E_{\text{g}}$ and the parity $\langle\hat{\Pi}\rangle$ are shown in the insets. (c) The von Neumann entropy $S_{\rm v-N}$ as well as the maximum of the quantum fluctuation $QF_{\rm max}$ is plotted along with the total number of excitations $\langle \hat{N}_{\rm ex} \rangle$ in the inset. (d) The average displacement coefficients $\overline f_k$ and $\overline g_k$ are shown for $\alpha = 0.01, 0.02, 0.03$, and $0.10$ (from top to bottom).  }
\label{f5}
\end{figure*}

\begin{figure*}[htbp]
\centering
\epsfysize=12cm \epsfclipoff \fboxsep=0pt
\setlength{\unitlength}{1.cm}
\begin{picture}(13,11)(0,0)
\put(0.0,0.0){{\epsffile{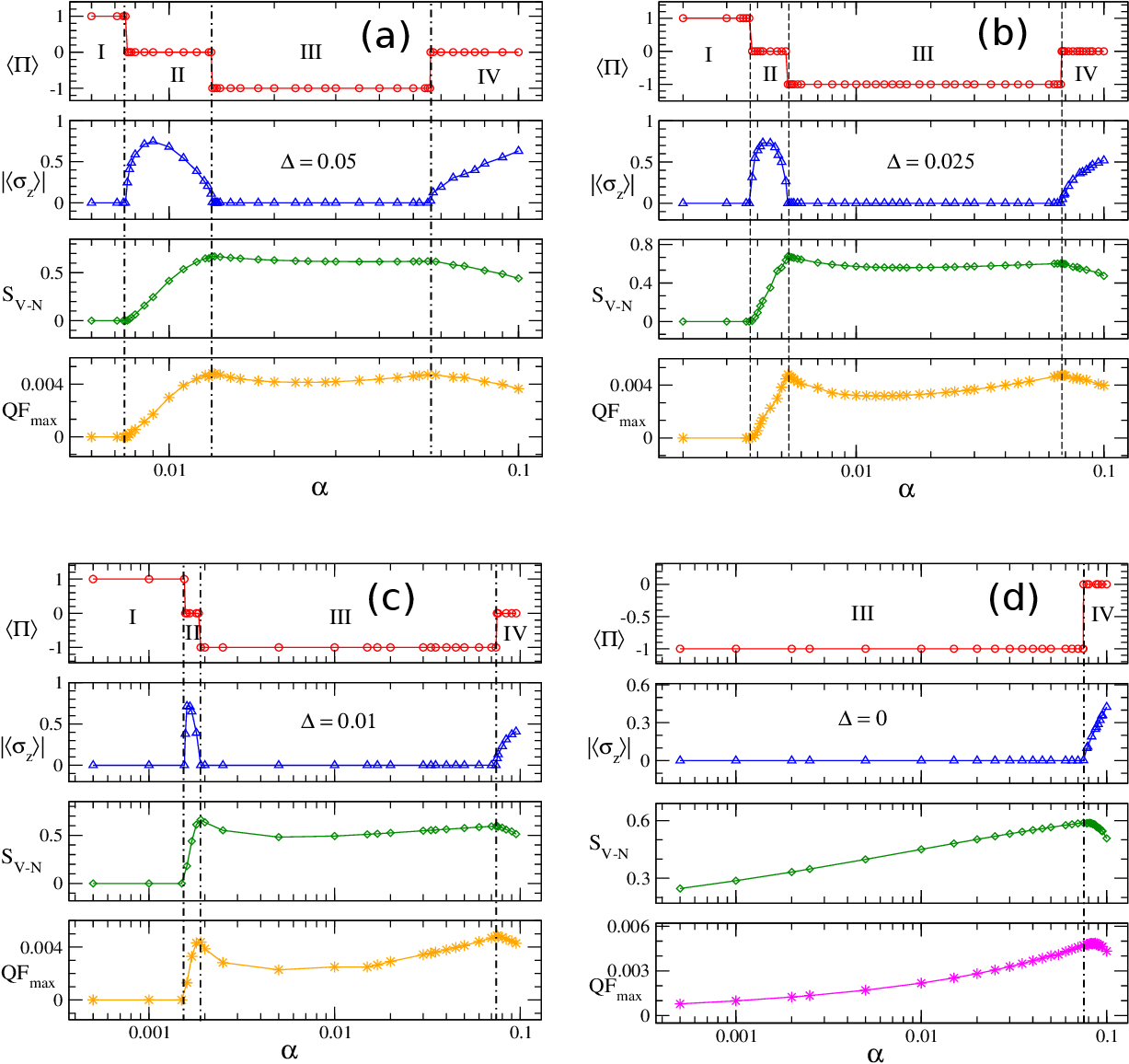}}}
\end{picture}
\caption{ For the weaker and vanishing tunnelings ($\Delta=0.05,0.025,0.01$, and $0$), ground-state properties including the parity $\langle\hat{\Pi}\rangle$, the spin magnetization $|\langle \sigma_z \rangle|$,  the von Neumann entropy $S_{\rm v-N}$, and the maximum of the quantum fluctuation $QF_{\rm max}$ (from top to bottom) are displayed as a function of the coupling $\alpha$ on a semi-log scale. The states I, II, III and IV are classified, and the transitions points are located by the dash-dotted lines. }
\label{f6}
\end{figure*}

Finally, the structure of ground-state wave function is carefully examined to better understand the critical properties. The logarithmic value of the ratio between average coherent-state weights $\log_{10}(\overline{A}/\overline{B})$ is plotted in Fig.~\ref{f3}(a) as a function of $\alpha$ for different $\Delta$. It vanishes in the delocalized phase, indicating the symmetry relation $\overline{A}=\overline{B}$. In the localized phase, however, $\log_{10}(\overline{A}/\overline{B})$ has two opposite nonzero values at the same coupling $\alpha$, corresponding to doubly degenerate ground states. For average displacement coefficients $\overline{f}_k$ and $\overline{g}_k$, a perfect antisymmetry (symmetry) relation is found over the whole range of frequencies $\omega_k$ in the upper (lower) panel of the Fig.~\ref{f3}(b), which is spontaneously broken at the transition point as shown in the middle panel. For further comparison, the results from the optimal displacement formula $\pm \lambda_k/(\omega_k+\chi)$ are also plotted with dashed lines. They provide a good fit to the numerical data.

In addition, the ground-state properties of the model in the off-diagonal coupling regime are investigated, where the coupling parameters satisfy $\lambda_k = -\gamma_k = \eta_k / 2$, using a tunneling amplitude $\Delta = 0.05$ as a representative example. Fig.~\ref{f4}(a) displays the variance of the ground-state energy, $\delta E_g^{(2)}(\alpha)$, as defined in Eq.~\eqref{delta_energy}. The transition point is identified as $\alpha_c = 0.0210(3)$, determined from the position of the peak. It is in excellent agreement with that obtained in the diagonal coupling case. Moreover, the entire curve remains below the threshold of $10^{-6}$, consistent with the result for the diagonal coupling case shown in Fig.~\ref{app_f2}(c), thereby reinforcing the accuracy and reliability of the variational approach adopted in this study. As shown in Fig.~\ref{f4}(b), in the localized phase ($\alpha > \alpha_c$), the expectation value of the momentum operator becomes non-zero, $\langle \hat{p}_k \rangle \neq 0$, while the position expectation value remains vanishing, $\langle \hat{x}_k \rangle = 0$. This behavior stands in stark contrast to the diagonal coupling scenario wherein $\langle \hat{x}_k \rangle \neq 0$ and $\langle \hat{p}_k \rangle = 0$ in the localized phase.

Further analysis reveals that the quantum criticality in the off-diagonal and diagonal coupling cases is equivalent, if one performs a simultaneous transformation of spin and bosonic operators: specifically, an exchange of spin magnetization components ($\sigma_z \rightarrow -\sigma_y$, $\sigma_y \rightarrow \sigma_z$) together with a swap of position and momentum operators ($\hat{x}_k \leftrightarrow -\hat{p}_k$). This equivalence originates from the underlying symmetry of the system.
Indeed, the SBM Hamiltonian in the off-diagonal coupling case can be mapped onto the diagonal form via a unitary rotation generated by $\hat{P}(\theta) = \exp(-i\theta \sigma_x / 2)$ \cite{jus25}.
For a rotation angle $\theta = -\pi/2$ and zero energy bias ($\varepsilon = 0$), the transformed Hamiltonian reads
\begin{eqnarray}
\hat{H}_{\rm rot}  &=& \hat{P}(\theta)^{-1}\hat{H}_{\rm offdiag}\hat{P}(\theta)  \nonumber \\
                   &=& -\frac{\Delta}{2}\sigma_x + \sum_{k}\omega_k b_{k}^\dag b_{k} + \frac{i\sigma_z}{2}\sum_k(\gamma_k-\lambda_k)(b^\dag_{k}- b_{k}) \nonumber \\
                   &=&  -\frac{\Delta}{2}\sigma_x + \sum_{k}\omega_k b_{k}^\dag b_{k} - \sigma_z\frac{\sqrt{2}}{2}\sum_k \eta_k\hat{p}_k.
\end{eqnarray}
It is evident that this rotated Hamiltonian assumes the same functional form as the standard diagonal-coupling model, thereby guaranteeing identical quantum critical behavior. The essential distinction lies in the nature of bath localization: in the diagonal case, localization occurs in position space ($\langle \hat{x}_k \rangle \neq 0$), whereas in the off-diagonal (or rotated) case, it manifests in momentum space ($\langle \hat{p}_k \rangle \neq 0$). Figs.~\ref{f4}(c) and (d) present a comprehensive comparison across the three models (diagonal, off-diagonal, and rotated), using multiple observables including the ground-state energy, spin magnetization, spin coherence, von Neumann entropy, quantum fluctuations, and symmetry parameters.
All results consistently confirm the equivalence of quantum criticality between the off-diagonal and diagonal coupling scenarios under the aforementioned transformations.

\subsection{RW coupling or CRW coupling}
\begin{figure*}[htb]
\centering
\epsfysize=6cm \epsfclipoff \fboxsep=0pt
\setlength{\unitlength}{1.cm}
\begin{picture}(15,6)(0,0)
\put(0.0,0.0){{\epsffile{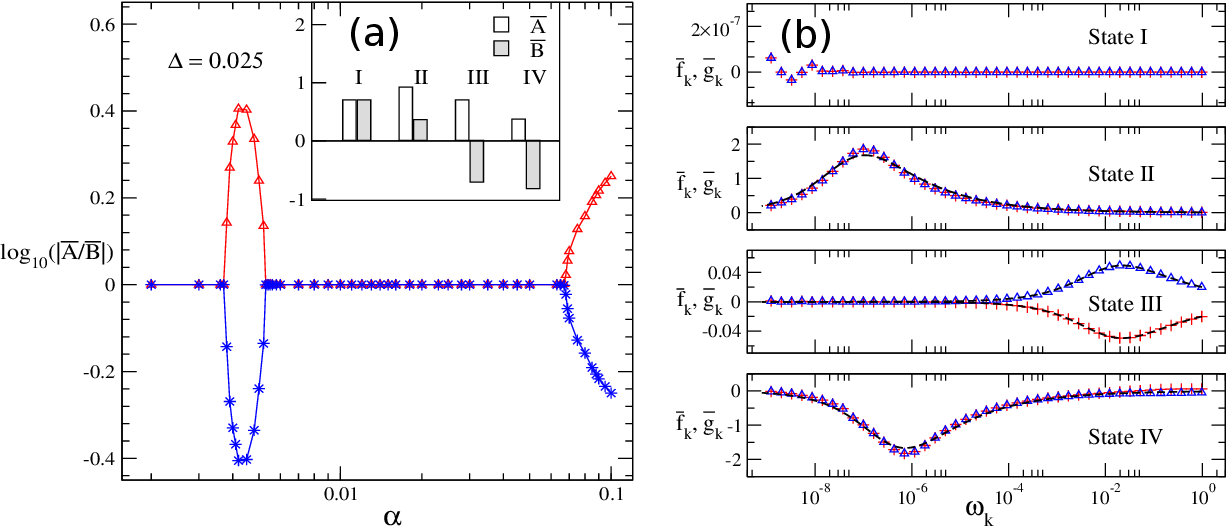}}}
\end{picture}
\caption{Taking the tunneling constant $\Delta=0.025$ as an example, the structure of the ground-state wave function is demonstrated with the average coherent-state weights $\overline{A}$ and $\overline{B}$, and average displacement coefficients $\overline{f}_k$ and $\overline{g}_k$  for the states I, II, III, and IV on a semi-log scale. The line with triangles/stars in (a) corresponds to one branch of the twofold degenerate states, and dashed lines in (b) stand for the fits with the optimal displacement formula. }
\label{f7}
\end{figure*}

\begin{figure*}[htb]
\centering
\epsfysize=11cm \epsfclipoff \fboxsep=0pt
\setlength{\unitlength}{1.cm}
\begin{picture}(13,11)(0,0)
\put(0.0,0.0){{\epsffile{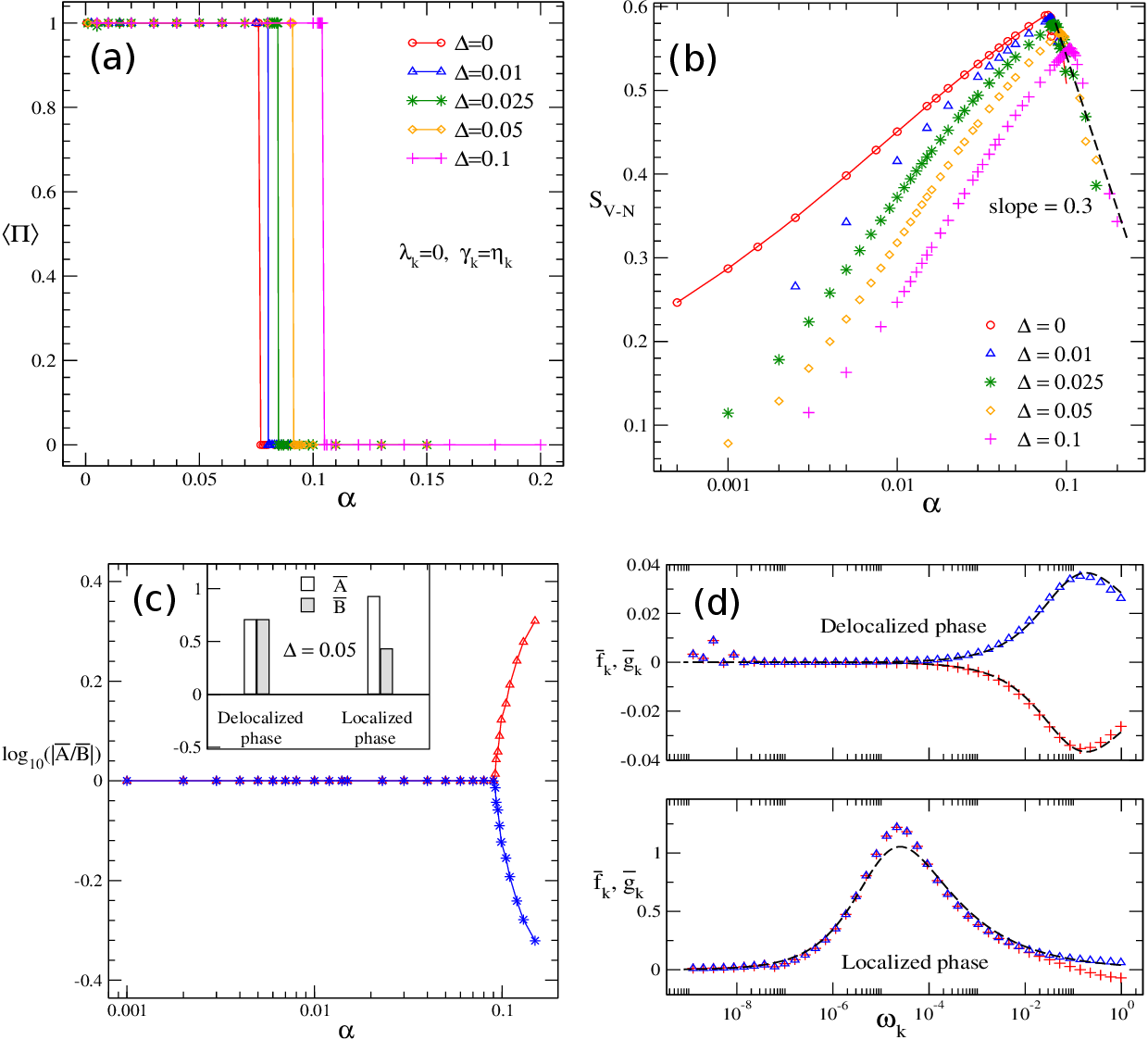}}}
\end{picture}
\caption{In the CRW coupling case, the parity $\langle \Pi \rangle$ in (a) and the von Neumann entropy $S_{\rm v-N}$ in (b) are plotted as a function of $\alpha$ for different $\Delta$. The dashed line represents a power-law fit. Average coherent-state weights $\overline{A}$ and $\overline{B}$ in (c) and average displacement coefficients $\overline{f}_k$ and $\overline{g}_k$ in (d) are demonstrated for the structural feature of the wave function at $\Delta=0.05$. The top and bottom panels correspond to the delocalized and localized phases, respectively. }
\label{f8}
\end{figure*}

To further explore the quantum criticality, the model under the RW coupling ($\lambda_k > 0, \gamma_k=0$) or CRW coupling ($\lambda_k=0, \gamma_k>0$) is investigated in this subsection. The strengths of the diagonal coupling $(\lambda_k + \gamma_k)/2 =\eta_k/2$ and off-diagonal coupling $(\lambda_k - \gamma_k)/2 = \pm \eta_k/2$ are then taken. At the strong tunneling $\Delta=0.1$, our variational results named as ``NVM$2$'' are plotted in Fig.~\ref{f5}(a) and (b) for the spin magnetization $|\langle \sigma_z \rangle|$ and spin coherence $\langle \sigma_x \rangle$ as well as the symmetry parameter $\langle\hat{\Pi}\rangle$. For comparison, earlier variational results in Ref.~\cite{wan19b} named as ``NVM$1$'' are also presented in which the logarithmic discretization factor $\Lambda=2$ is used, much greater than the requirement of the continuum limit $\Lambda \rightarrow 1$. The ground-state energy $E_g$ of NVM$2$ is much lower than that of NVM$1$ in the strong $\alpha$ regime, showing that the bath with a high dense spectrum is demanded to obtain an accurate description of quantum transitions. Moreover, our variational results deviate significantly from those of NVM$1$ especially for the spin magnetization, but agree well with numerical results obtained by VMPS, demonstrating the validity of NVM calculations in this work.

In Fig.~\ref{f5}(c), the von Neumann entropy $S_{\rm v-N}$ and quantum fluctuation $QF_{\rm max}$ are presented for further investigation. A QPT is identified at $\alpha_c=0.015$ by the emergence of the nonzero $S_{\rm v-N}$ and $QF_{\rm max}$ in conjunction with the behavior of $|\langle \sigma_z \rangle|$. For the coupling strength $\alpha < \alpha_c$, the total excitations number $\langle \hat{N}_{\rm ex}\rangle \approx 10^{-9}$ is derived in the inset, eight orders of magnitude smaller than that of the diagonal coupling case. The results indicate that the excitation entirely vanishes in the RW coupling case. Consequently, the system possesses a higher $U(1)$ symmetry. Taking $\alpha=0.01$ as an example, the result $\overline{f}_k =\overline{g}_k=0$ is obtained in the top panel of Fig.~\ref{f5}(d), suggesting that the bosonic bath is in the vacuum state. The anisotropic model is thus in the so-called ``free'' phase, as the impurity retains the characteristics of a free spin even in the presence of a nonzero coupling. Further studies give the wave function of the ground state $|\Psi_g \rangle = \frac{\sqrt{2}}{2}(| \uparrow \rangle + | \downarrow \rangle) \otimes |0\rangle_{\textrm{b}}$.

In the strong coupling regime $\alpha > \alpha_c$, a non-monotonic $\alpha$-dependence of the order parameter $|\langle \sigma_z \rangle|$ is observed, quite different from that of the diagonal coupling case in Fig.~\ref{f1}(b). Three points with $\alpha=0.02,0.03$, and $0.10$ are selected as illustrative examples to examine the structure of the wave function, as depicted in the lower panels of Fig.~\ref{f5}(d). The symmetry relation $\overline{f}_k =\overline{g}_k$ is found over the whole range of frequencies for all the three cases, indicating that all of them are in the localized phase. In the inset of the subfigure (c), the number of total excitations $\langle \hat{N}_{\rm ex}\rangle$ fluctuates around the number of the bath modes $M=430$. It differs significantly from that of the diagonal coupling case where $\langle \hat{N}_{\rm ex}\rangle$ increases monotonically with the coupling $\alpha$. Besides, our finding contradicts the claim in Refs.~\cite{ton11,wan19b,wan20} that $\langle \hat{N}_{\rm ex}\rangle$ in the SBM under RW approximation exhibits discrete jumps between integer-valued plateaus, yielding level-crossing-induced transition. This demonstrates remarkable superiority of NVM in studying QPTs.

In addition, the spin coherence $\langle \sigma_x \rangle$ consistently remains positive in the diagonal coupling scenario at $\Delta=0.1$. By contrast, in the case of RW coupling,  $\langle \sigma_x \rangle$ rapidly declines to a negative value and then gradually approaches zero as $\alpha$ increases. It indicates that the off-diagonal system-bath interaction supersedes the tunneling effect $\Delta$ and becomes the dominant factor within the strong coupling regime. Further analysis on the rounded summit of the curves at $\alpha \approx 0.04$ for $S_{\rm v-N}$ and $QF_{\rm max}$ suggests that a smooth crossover rather than QPT takes place.  Thus, the localized phase can be divided into two parts with different signs of the spin coherence $\langle \sigma_x \rangle$, resulting in the abnormal behavior of $|\langle \sigma_z \rangle|$.

For weaker tunnelings $\Delta=0.05,0.025$, and $0.01$, four different sates I, II, III, and IV are classified according to the behaviors of $\langle\hat{\Pi}\rangle$,  $|\langle \sigma_z \rangle|$,  $S_{\rm v-N}$ and $QF_{\rm max}$. They are separated by three QPTs, and the values of critical couplings $\alpha_{c1}, \alpha_{c2}$, and $\alpha_{c3}$ (from left to right) are denoted by the dash-dotted lines, as shown in Fig.~\ref{f6}(a)-(c). As the tunneling $\Delta$ decreases, both $\alpha_{c1}$ and $\alpha_{c2}$ exhibit a rapid decrease, and ultimately vanish at $\Delta=0$. This behavior is remarkably different from that of $\alpha_{c3}$. In Fig.~\ref{f6}(d), only a single QPT is found between states III and IV.

Setting $\Delta=0.025$ as an example, further investigation on these four states is performed by means of the average coherent-state weights $\overline{A}$ and $\overline{B}$ in Fig.~\ref{f7}(a) and average displacement coefficients $\overline{f}_k$ and $\overline{g}_k$ in Fig.~\ref{f7}(b), respectively. Similarly, the results $\overline{A}=\overline{B}=\frac{\sqrt{2}}{2}$ and $\overline{f}_k =\overline{g}_k=0$ indicate that state I belongs to the free phase. Both states II and IV are in the localized phase based on the relation $\overline{f}_k =\overline{g}$, though the signs of the ratio $\overline{A}/\overline{B}$ are opposite, as shown in the inset of the subfigure (a). It means that the spin coherence $\langle \sigma_x \rangle$ is positive (negative) in the state II (IV), in agreement with that in Fig.~\ref{f5}(b). Besides, the antisymmetry relation $\overline {f}_k=-\overline {g}_k$ is found in the state III, suggesting that it belongs to the delocalized phase. The relation $\overline{A}=-\overline{B}$ is in contrast to that in the diagonal coupling case, yielding a new parity $\langle\hat{\Pi}\rangle = -1$ as depicted in Fig.~\ref{f6}. In fact, the standard SBM in the weak coupling regime also has such odd parity when the value of the tunneling $\Delta$ is set to be negative. Thus, the odd-parity state III may arise from the negative effective tunneling $\Delta_{\rm eff}$.

\begin{figure*}[htb]
\centering
\epsfysize=6cm \epsfclipoff \fboxsep=0pt
\setlength{\unitlength}{1.cm}
\begin{picture}(13,6)(0,0)
\put(0.0,0.0){{\epsffile{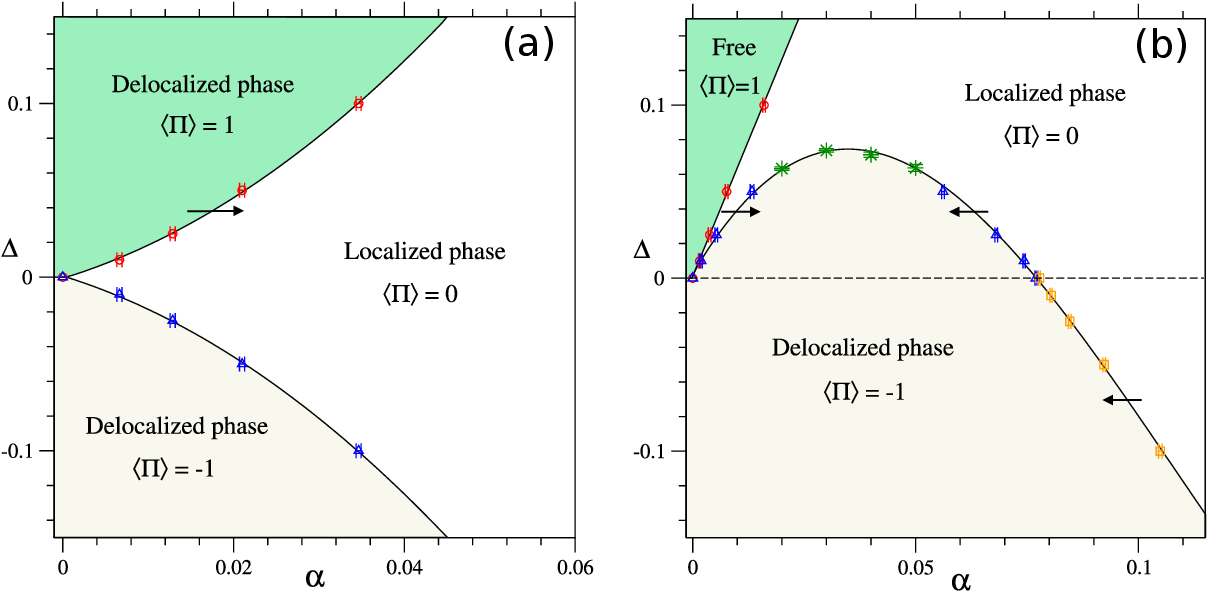}}}
\end{picture}
\caption{Phase diagram of the ASBM is displayed in the $\alpha-\Delta$ plane for the diagonal coupling case in (a) and RW coupling case in (b). Both the free phase and delocalized phase have conserved $Z_2$ symmetry $\langle\hat{\Pi}\rangle=\pm 1$, and the symmetry is spontaneously broken in the localized phase with $\langle\hat{\Pi}\rangle=0$. The phase boundary is depicted by the circles and triangles representing critical coupling $\alpha_c(\Delta)$ as well as the stars  $\Delta_c(\alpha)$ in (b). Squares are obtained from the CRW coupling case by using the relation $\alpha_c^{\rm RW}(\Delta) = \alpha_c^{\rm CRW}(-\Delta)$. Solid lines show linear/polynomial fits, and arrows indicate localized-to-delocalized  transitions.  }
\label{f9}
\end{figure*}

\begin{figure*}[htb]
\centering
\epsfysize=12cm \epsfclipoff \fboxsep=0pt
\setlength{\unitlength}{1.cm}
\begin{picture}(13,11)(0,0)
\put(0.0,0.0){{\epsffile{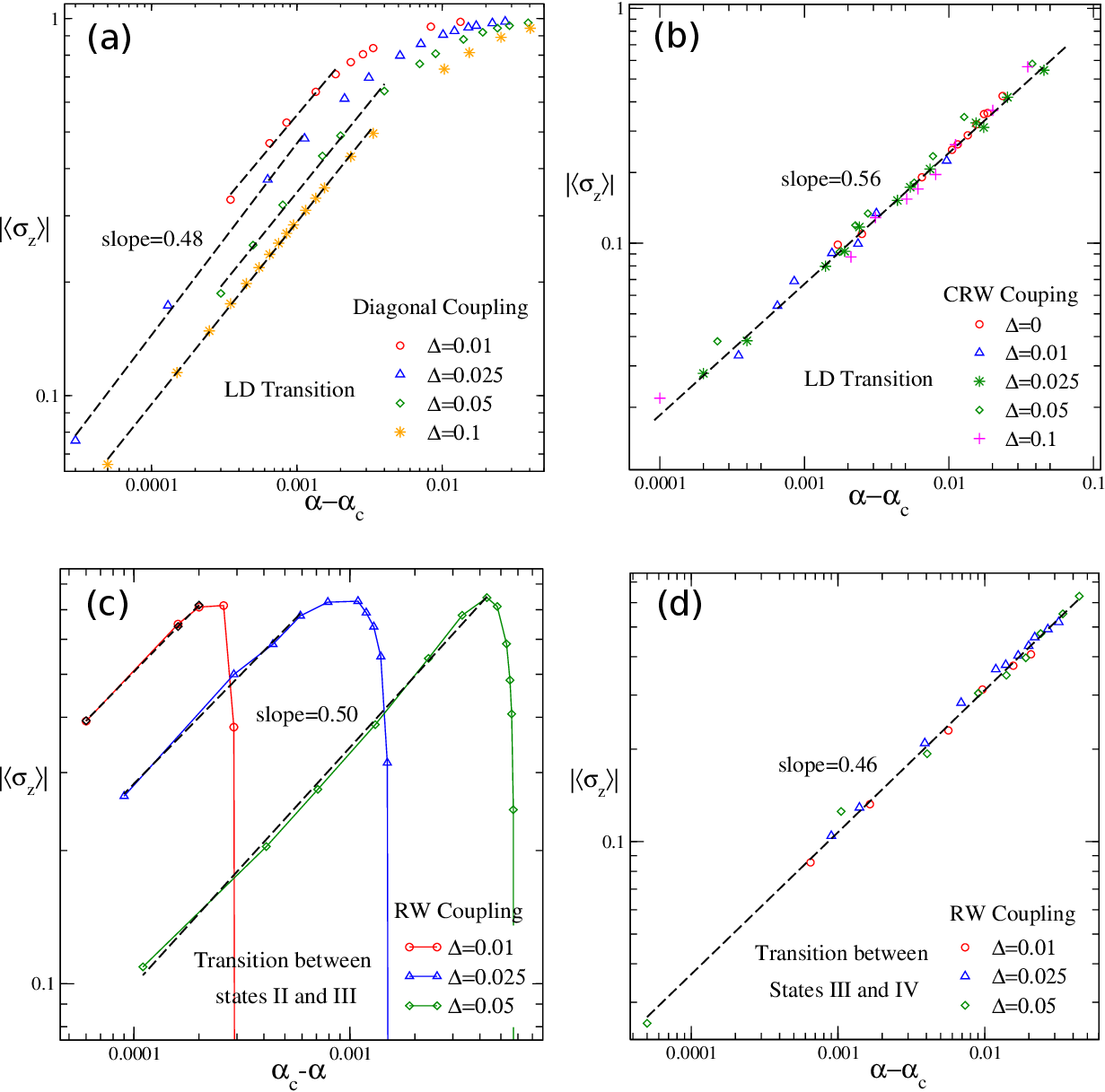}}}
\end{picture}
\caption{ The spin magnetization $|\langle\sigma_z\rangle|$ is displayed as a function of the shift $|\alpha-\alpha_c|$ on a log-log scale for different tunneling amplitude $\Delta$ in the diagonal coupling case (a), the CRW coupling case (b), and the RW coupling case (c) and (d). Dashed lines represent power-law fits. }
\label{f10}
\end{figure*}

For comparison, the ASBM under the CRW coupling is investigated, and the results are shown in Fig.~\ref{f8}. For each specified tunneling parameter $\Delta=0,0.01,0.025,0.05$, and $0.1$, a single QPT is discerned through the symmetry parameter $\langle\hat{\Pi}\rangle$ and the von Neumann entropy $S_{\rm v-N}$. This scenario mirrors that of the case of RW coupling at the tunneling $\Delta=0$ as given in Fig.~\ref{f6}(d). The critical coupling strength exhibits a monotonic increase with the augmentation of the tunneling parameter. Specifically, $\alpha_c=0.076(1)$ is determined at the vanishing tunneling $\Delta$, which aligns with that between the states III and IV. Besides, a cusp is observed at each bifurcation point for the von Neumann entropy $S_{\rm v-N}$, suggesting that the transition is of second order. Remarkably, all the data converge to a single curve following $\alpha_c$ exhibiting a consistent slope of $0.3$, showing the $\Delta$-independence. Taking $\Delta=0.025$ as a representative case, the analysis on the ground-state wave function, as depicted in Fig.~\ref{f8}(c) and (d), corroborates that the QPT is indeed a localized-delocalized transition, although it arises from the interplay between the diagonal and off-diagonal couplings.

\section{Discussion}\label{sec:dis}

In this study, a comprehensive investigation of the ground-state properties of the ASBM within a unified framework is carried out, both with and without the rotating-wave approximation. Through systematic comparison and analysis of various observables, several distinct quantum states are identified, categorized into the delocalized phase with even/odd parity, the localized phase, and the free phase. Notably, quantum phase transitions between these states are accompanied by spontaneous symmetry breaking of either discrete $Z_2$ or continuous $U(1)$ symmetry. It suggests that these observed transitions are continuous in nature. A comprehensive phase diagram is subsequently constructed in the parameter space spanned by tunneling strength and spin-bath coupling. The critical boundaries for all transitions are precisely measured and summarized in Fig.~\ref{f9}, where the subfigure (a) corresponds to the diagonal coupling case, and the subfigure (b) to the RW coupling case.

To better understand the phase diagram of the ASBM, the Hamiltonian expectation expression is derived by substituting the trial wave function $|\Psi_g\rangle$ of the ground state defined in Eq.~(\ref{vmwave1}),
\begin{equation}
\langle\Psi_g|\hat{H}|\Psi_g\rangle   =    H_{\rm b} + H_{\rm diag} + H_{\rm offdiag} + H_{\rm tunnel},
\end{equation}
where $H_{\rm tunnel}, H_{\rm diag}, \,{\rm and}\, H_{\rm off-diag}$ denote energy terms for the tunneling, diagonal and off-diagonal couplings,
respectively, and $H_{\rm b}$ is related to the bosonic bath,

\begin{eqnarray}
\label{Hamilton3}
H_{\rm b} & = &\sum_{m,n} A^{*}_mA_nF_{m,n}\sum_k\omega_kf^{*}_{m,k}f_{n,k} \nonumber \\
& + & \sum_{m,n} B^{*}_mB_nG_{m,n}\sum_k\omega_kg^{*}_{m,k}g_{n,k},
\end{eqnarray}
\begin{eqnarray}
H_{\rm diag} & = & \sum_{m,n} A^{*}_mA_nF_{m,n}\sum_k\frac{\lambda_k+\gamma_k}{2}(f^{*}_{m,k}+f_{n,k}) \nonumber \\
& - & \sum_{m,n} B^{*}_mB_nG_{m,n}\sum_k\frac{\lambda_k+\gamma_k}{2}(g^{*}_{m,k}+g_{n,k}), \nonumber \\
H_{\rm offdiag} & = & \sum_{m,n} A^{*}_mB_n\Gamma_{m,n}\sum_k\frac{\gamma_k -\lambda_k}{2}(f^{*}_{m,k} - g_{n,k}) \nonumber \\
                & + &  \sum_{m,n} B^{*}_mA_nK_{m,n}\sum_k\frac{\gamma_k -\lambda_k}{2}(f_{n,k} - g^{*}_{m,k}), \nonumber \\
H_{\rm tunnel} & = & - \frac{\Delta}{2}\sum_{m,n} \left( A^{*}_mB_n\Gamma_{m,n} + B^{*}_mA_nK_{m,n} \right) .
\end{eqnarray}
As mentioned before, the emergence of rich phase transitions  arises from the competition among $H_{\rm diag}, H_{\rm offdiag}$ and $H_{\rm tunnel}$.

In the diagonal coupling case with $\lambda_k=\gamma_k$, the off-diagonal coupling term $H_{\rm offdiag}$  becomes identically vanishing. Thus, a sharp QPT demarcates the boundary between localized and delocalized quantum phases. Under positive tunneling  $\Delta >0$, the self-consistency equations in Eq.~(\ref{vmwave1}) concerning $H_{\rm tunnel}$ require $A_m$ and $B_n$ coefficients to share identical algebraic signs for ground-state realization. Consequently, the spin coherence  $\langle \sigma_x \rangle = \sum_{m,n}A^{*}_mB_n\Gamma_{m,n} + B^{*}_mA_nK_{m,n}$ manifests a positive value, aligning with prior numerical results. By the definition of the parity operator $\hat{\Pi}=\sigma_x\exp\left[i\pi\sum_k b^\dag_{k}b_k\right]$, the symmetry parameter exhibits strong correlation with the spin coherence. Specifically, even parity $\langle\hat{\Pi}\rangle = 1$ emerges in the delocalized phase through this operator's eigenvalue structure, as confirmed by the numerical results in Fig.~\ref{f1}(a). Remarkably,  the ground-state energy remains invariant under tunneling sign inversion $\Delta \to -\Delta$, which can be achieved through the transformation $A_m \to -A_m$ or  $B_n \to -B_n$. Such symmetry operation inverts both the spin coherence and parity, while preserving the critical coupling strength relationship $\alpha_c(\Delta)=\alpha_c(-\Delta)$.  The resultant phase diagram therefore exhibits a mirror symmetry about $\Delta = 0$.

Owing to the interplay of three competing terms, the phase diagram becomes richer in the RW coupling scenario wherein $\lambda_k>0$ and $\gamma_k=0$, as illustrated in Fig.~\ref{f9}(b). Three distinct phases are identified by the symmetry parameter $\langle \hat{\Pi}\rangle$, which are the free phase (even parity), the delocalized phase (odd parity), and the localized phase (parity breaking). Compared to the diagonal coupling case, the delocalized phase with even parity appears to be largely replaced by the free phase due to the presence of the off-diagonal coupling term $H_{\rm offdiag}$. Further analysis reveals that the free phase exhibits higher $U(1)$ symmetry. The phase boundaries are delineated by circles and triangles representing the critical coupling $\alpha_c(\Delta)$, as well as by stars denoting $\Delta_c(\alpha)$.

Additionally, the ground-state energy exhibits invariance under the transformations $\Delta \to -\Delta$, $B_n \to -B_n$, and the switch from the case involving RW coupling  to that involving CRW coupling wherein the strength of the off-diagonal coupling is $(\gamma_k -\lambda_k)= \pm \eta_k$. This invariance establishes the relationship $\alpha_c^{\rm RW}(\Delta) = \alpha_c^{\rm CRW}(-\Delta)$, indicating that the phase diagram of the RW-coupled system, despite its inherent asymmetry, exhibits exact mirror symmetry about $\Delta=0$ relative to its CRW-coupled counterpart. To confirm the prediction, numerical results for the phase boundary of the CRW-coupled case ($\Delta \ge 0$) are plotted with squares after a mirror flip. The curve smoothly merges with that of the RW-coupled case at the point $\alpha_c(\Delta=0) = 0.076(1)$, affirming the existence of mirror symmetry between the two coupling scenarios.

The special value of the tunneling  $\Delta^*=0.074(1)$, as the maximum of $\Delta_c(\alpha)$, is detected via third-order polynomial fitting (solid line) for the entire phase boundary between the localized and delocalized phases. When the tunneling strength exceeds this critical threshold, the system undergoes a distinct phase transition, referred to as the free-localized transition. Subsequent analysis demonstrates that the transition is characterized by second-order criticality, accompanied by spontaneous symmetry breaking in the continuous $U(1)$ gauge group \cite{bak14,hwa16}. For a weaker tunneling $0< \Delta < \Delta^*$, the pathway of the phase transition becomes extremely complicated. Successive transitions are observed through the free phase (state I), localized phase (state II), odd-parity delocalized phase (state III), and ultimately localized phase (state IV), establishing a multi-stage phase transition sequence. Furthermore, the localized phases II and IV can be distinguished by the spin coherence, corresponding to identical/opposite signs in the averaged superposition weights of coherent states, as illustrated in the inset of Fig.~\ref{f7}(a). In the case of negative tunneling $\Delta < 0$, the system returns to the scenario of a single QPT.

In fact, four different localized-delocalized transitions, marked by the arrows, are found in the phase diagrams of ASBM, corresponding to the cases of diagonal coupling, CRW coupling, and RW coupling, respectively. As shown in Fig.~\ref{f10}, the order parameter $|\langle \sigma_z \rangle|$ is plotted as a function of the coupling shift $\tau = |\alpha-\alpha_c|$ for different values of the tunneling $\Delta$ on a log-log scale. Power-law behaviors of $|\langle\sigma_z\rangle|$ are expected with respect to $\tau$, and the values of the critical exponent $\beta$ can be measured from the slopes. In the subfigures (a) and (c), the growing curves are almost parallel to each other, showing the significant $\Delta$ dependency, and the results $\beta=0.48(1)$ and $0.50(1)$ are quite close to the mean-field prediction $\beta=1/2$. It indicates that the transition between the states II and III is identical to that in the conventional SBM, and the off-diagonal coupling term $H_{\rm offdiag}$ is then insignificant. In contrast, numerical data of $|\langle \sigma_z \rangle|(\tau)$ for different $\Delta$ collapses onto a single curve in both the subfigures (b) and (d), corresponding to the cases of CRW coupling and strong RW coupling, respectively. It implies that the phase transitions stem from the interplay between diagonal and off-diagonal couplings, irrelevant to the tunneling $\Delta$. The measured slopes of $0.56(2)$ and $0.46(2)$ slightly deviations from the mean-field prediction $1/2$. It may be attributed to the influence of the off-diagonal coupling.

Critical behaviors of higher-order quantum observables, such as the Kondo energy scale $\chi$, have not been involved in this work due to limitations in the sophistication of the variational ansatz. Recently, however, a new implementation of the coherent expansion was proposed based on analytical treatment \cite{flo15}. It was demonstrated that the momentum-dependent variational parameters $f_{n,k}$ and $g_{n,k}$ can be expressed as rational functions of the frequency $\omega_k$. This crucial property enables the recasting of the optimization problem into a finite-dimensional parameter space, consisting of the coefficients of these rational functions and the weights of the coherent states. Consequently, the total number of variational parameters is reduced to $N^2+N-1$, which is independent of the bath discretization size $M$. This technical advancement may allow for the systematic exploration of QPTs in the thermodynamic limit using an extremely large coherent-state number $N$, with the Kondo energy scale $\chi$ included.

\section{Conclusion}\label{sec:con}

Through large-scale numerical variational calculations, quantum phase transitions in the anisotropic spin-boson model with three competing terms are systematically investigated, employing a sub-Ohmic bath characterized by high spectral density. Four cases involving distinct system-environment interactions, which are diagonal, off-diagonal, RW, and CRW, are carefully examined. Comprehensive analyses of spin-related observables, symmetry parameter, quantum entanglement and fluctuation, and wave-function structure are performed, enabling precise determination of transition points and critical exponents across varying tunneling strengths. Additionally, a rich phase diagram is revealed in the parameter space of tunneling and spin-bath coupling, with detailed comparisons made to its prototype model. Our results demonstrate excellent agreement with those obtained by VMPS, thereby lending support to the validity of the variational calculations in this work.

Remarkably, even at a low spectral exponent (e.g., $s=0.3$), a complex phase transition landscape is observed in the RW-coupled case, contrasting prior findings \cite{wan20}. Beyond conventional delocalized (with conserved $Z_2$ symmetry) and localized (with broken $Z_2$ symmetry) phases, a novel phase exhibiting higher $U(1)$ symmetry is identified. Two second-order critical lines, corresponding to spontaneous $Z_2$ and $U(1)$ symmetry breaking, intersect at the origin, asymmetrically partitioning the phase space. Specifically, the system undergoes three QPTs with increasing coupling for weak tunneling $0 < \Delta  < \Delta^*=0.074(1)$: from the free phase (state I) to the localized phase (state II), then to the odd-parity delocalized phase (state III), and finally back to the localized phase (state IV). In contrast, only a single QPT occurs beyond this range. Notably, the phase diagram displays exact mirror symmetry relative to its CRW-coupled counterpart about $\Delta=0$. Mirror symmetry is also observed in the phase diagram of the diagonal coupling case, whereas the crossover takes place instead of the QPT in the off-diagonal coupling one.

The odd-parity delocalized phase is found in both anisotropic and prototype models, but the underlying mechanisms may be different. Critical behavior analysis reveals that all four localized–delocalized transitions exhibit mean-field critical exponents, despite differing origins: two transitions governed by the fundamental tunneling-dissipation competition, and two arising from the intricate interplay between diagonal and off-diagonal coupling terms. Given its potential realization in superconducting circuit QED systems, the anisotropic model provides a pivotal platform for exploring the complex quantum critical phenomena emerging from the competition of these fundamental interactions.

{\bf Acknowledgements:} This work was supported in part by National Natural Science Foundation of China (under Grant Nos.~$11875120$ and $12175052$), Hangzhou Joint Fund of the Zhejiang Provincial Natural Science Foundation of China (under Grant No. LHZSD24A050001), and Hangzhou Leading Youth Innovation and Entrepreneurship Team project (under Grant No. TD2024005).

The data that support the findings of this article are openly available \cite{zhou_data25}.

\appendix
\counterwithin{figure}{section}
\section{the benchmark of NVM}\label{sec:appendix}

\begin{figure*}[htb]
\centering
\epsfysize=6cm \epsfclipoff \fboxsep=0pt
\setlength{\unitlength}{1.cm}
\begin{picture}(13,6)(0,0)
\put(0.0,0.0){{\epsffile{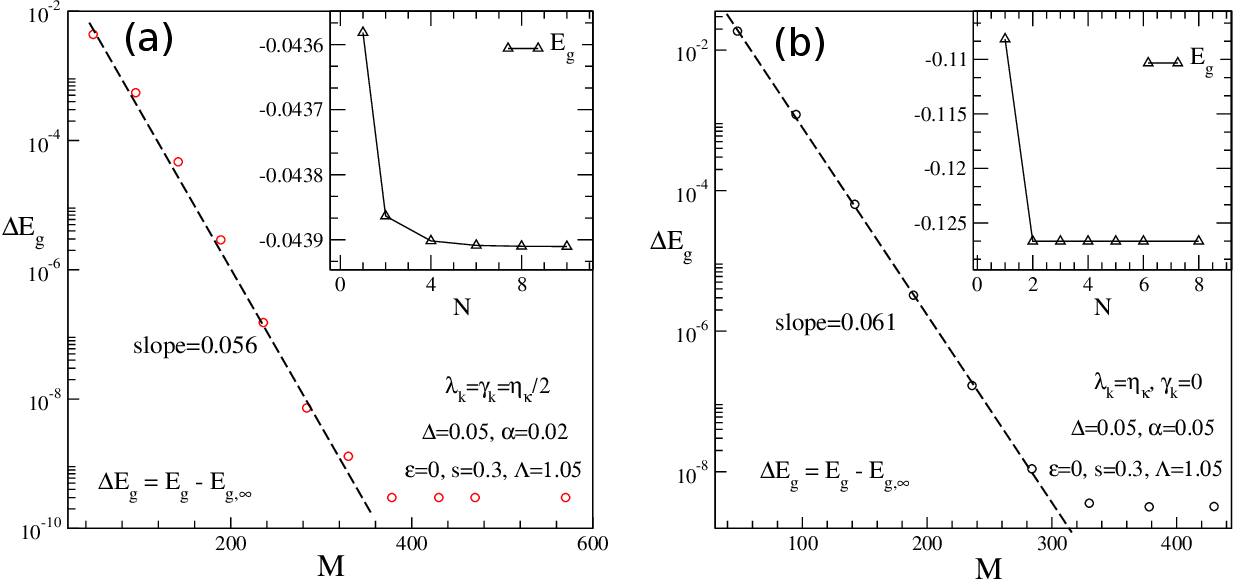}}}
\end{picture}
\caption{ The ground-state energy $E_g$ and the shift $\Delta E_g = E_g - E_{g,M=\infty}$ are displayed against the bath-mode number $M$ and coherent-state superposition number $N$ for the diagonal coupling case in (a) and RW coupling case in (b). Dashed lines represent the fits with an exponential form. For clarity, the curves of $\Delta E_g$ are shifted up by a tiny amount.
}
\vspace{2.5\baselineskip}
\label{app_f1}
\end{figure*}

The convergence of numerical variational results is carefully examined for the numbers of bath modes $M$ and coherent states $N$ in Fig.~\ref{app_f1} and its inset, respectively. The setting of the logarithmic-grid discretization factor $\Lambda=1.05$, spectral exponent $s=0.3$, tunneling amplitude $\Delta=0.05$, and energy bias $\varepsilon=0$ is adopted. The coupling strengths $\alpha=0.02$ and $0.05$, slightly smaller than the critical coupling $\alpha_c$, are fixed in subfigures (a) and (b), corresponding to the diagonal coupling and RW coupling cases, respectively, both of which are in the delocalized phase. As the bath-mode number $M$ increases, the shift of the ground-state energy $\Delta E_{\rm g} = E_{\rm g} - E_{\rm g, M=\infty}$ in both subfigures (a) and (b) decays exponentially. Here, $E_{\rm g, M=\infty}$ represents the asymptotic value of $E_{\rm g}(M)$, and the slopes of $0.056$ and $0.061$ are nearly identical. It indicates that $M=430$ is sufficiently large for the convergence of NVM calculations with the logarithmic grid at $\Lambda=1.05$ \cite{zho22}.

\begin{figure*}[htb]
\centering
\epsfysize=10cm \epsfclipoff \fboxsep=0pt
\setlength{\unitlength}{1.cm}
\begin{picture}(13,6)(0,0)
\put(1.0,0.0){{\epsffile{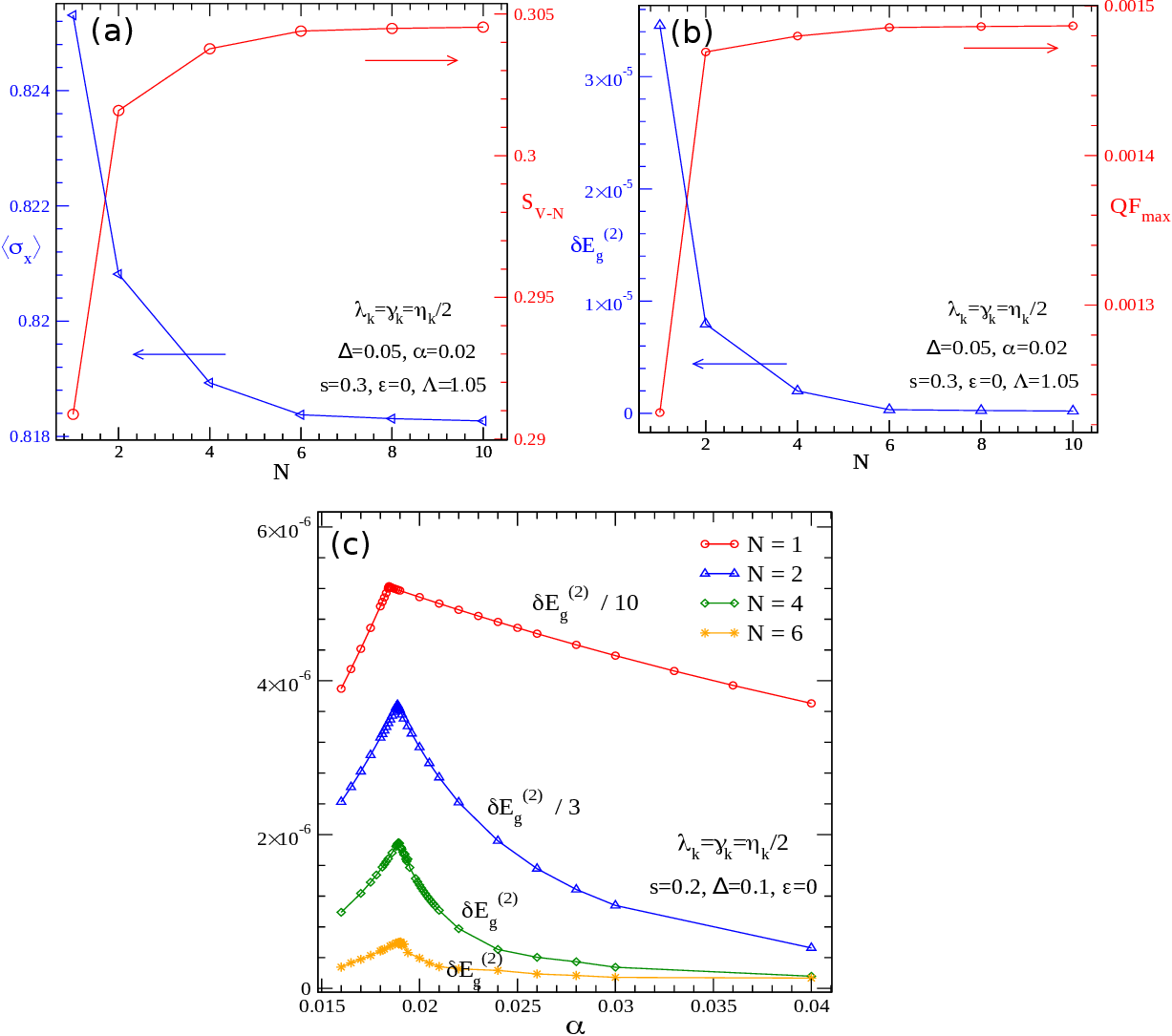}}}
\end{picture}
\caption{In the diagonal coupling case, the spin coherence $\langle \sigma_x \rangle$ and von Neumann entropy $S_{\rm v-N}$ in (a), and the variance of the ground-state energy $\delta E_g^{(2)}$ as well as the maximum of the quantum fluctuation $QF_{\rm max}$ in (b) are displayed with respect to the number of the multiplicity $N$, taking the coupling $\alpha=0.02$ as an example. (c) The variance of the energy $\delta E_g^{(2)}$ as a function of the coupling $\alpha$ is also plotted for different multiplicity $N=1,2,4$ and $6$. For clarity, the curves at $N=1$ and $2$ have been rescaled by the factors $1/10$ and $1/3$, respectively.
}
\label{app_f2}
\end{figure*}

In the insets of Fig.~\ref{app_f1}, the convergence of the ground-state energy $E_{\rm g}$ with respect to the number of coherent superposition states, $N$, is further examined. In contrast to the behavior of $E_{\rm g}(M)$, the energy $E_{\rm g}(N)$ exhibits a rapid decrease and converges much more swiftly to its asymptotic value, $E_{\rm g}(N=\infty)$. It suggests that a modest number of coherent states, namely, $N=6$ for the diagonal coupling case and $N=4$ for the RW coupling case, is sufficient to obtain reliable results. Furthermore, we have conducted additional convergence tests for other key observables, including the spin magnetization $|\langle\sigma_z\rangle|$, spin coherence $\langle\sigma_x\rangle$, von Neumann entropy $S_{\rm v-N}$, and maximum quantum fluctuation $QF_{\rm max}$. The results presented in Figs.~\ref{app_f2}(a) and (b) reveal that all these observables converge rapidly, providing strong support for our assertion that $N=6$ is sufficient for the diagonal coupling case.

To rigorously assess the validity of NVM, we also introduce the variance of the ground-state energy,
\begin{equation}
\label{delta_energy}
\delta E_{g}^{(2)}= \langle \Psi_{\rm g}|\hat{H}^2 |\Psi_{\rm g}\rangle - \langle \Psi_{\rm g}|\hat{H}|\Psi_{\rm g}\rangle^2,
\end{equation}
which serves as an important benchmark for the quality of NVM. Our calculations show that the variance $\delta E_{g}^{(2)}$ converges to a value on the order of $10^{-7}$, a result that attests to the validity of our approach.
In additional, the behavior of the variance was also systematically examined across the full parameter space of the coupling for various multiplicities $N=1,2,4$, and $6$. As illustrated in Fig.~\ref{app_f2}(c), each curve displays a pronounced peak at the critical point, the magnitude of which diminishes monotonically with increasing $N$. Notably, for $N=6$, the variance is maintained below the threshold of $10^{-6}$, which highlights the robustness and precision of our variational calculations.

\begin{figure*}[htb]
\centering
\epsfysize=6cm \epsfclipoff \fboxsep=0pt
\setlength{\unitlength}{1.cm}
\begin{picture}(13,6)(0,0)
\put(0.0,0.0){{\epsffile{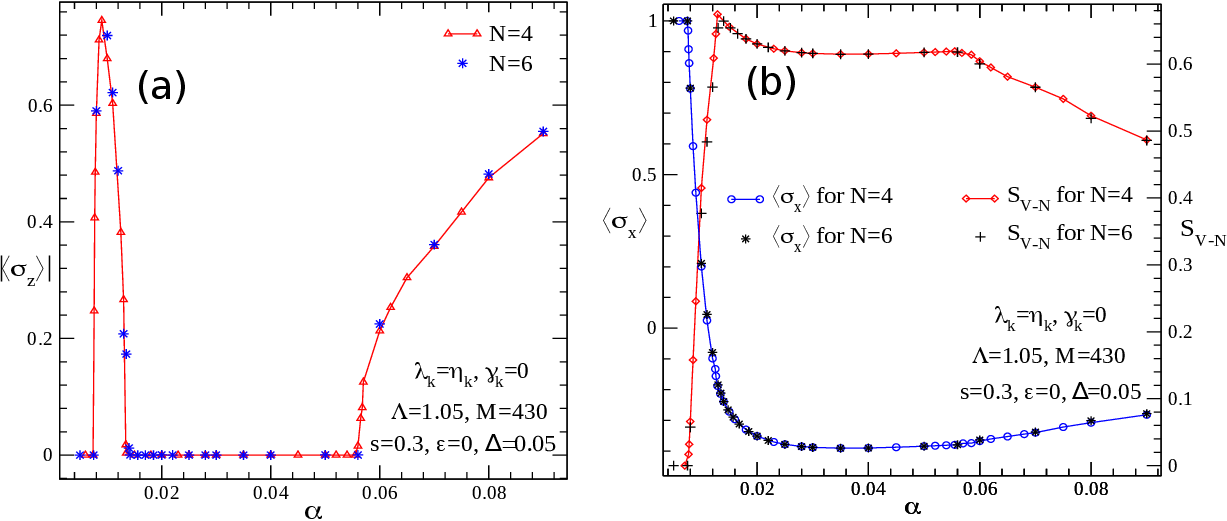}}}
\end{picture}
\caption{In the RW coupling case,  the coupling-dependent curves of the spin magnetization $|\langle \sigma_z \rangle|$ in (a),  spin coherence $\langle \sigma_x \rangle$, and the von Neumann entropy $S_{\rm v-N}$ in (b) are presented for
the number of the multiplicity $N=4$ (lines with dots) and $N=6$ (dots).
}
\label{app_f3}
\end{figure*}

For the RW coupling case, additional calculations with $N=6$ were also conducted in a comparative analysis, using a tunneling amplitude of $\Delta=0.05$ as an example. Figure~\ref{app_f3} demonstrates that the results obtained with $N=4$ and $N=6$ are in  a good agreement for the spin magnetization, spin coherence, and von Neumann entropy over the entire coupling regime. It provides further confirmation that a multiplicity of $N=4$ is fully sufficient for the precise characterization of phase transitions within the RW coupling framework.

\end{document}